\documentclass[aps,prd,onecolumn,fleqn,superscriptaddress,groupedaddress,nofootinbib,preprintnumbers]{revtex4}
\pdfoutput=1
\usepackage{amsmath,amssymb,pstricks,graphicx,todonotes,xspace}
\usepackage{todonotes}
\usepackage{subfig}
\usepackage{relsize}
\usepackage{booktabs}
\usepackage{tabularx}
\usepackage{bm}
\usepackage{verbatim}
\usepackage{multirow}
\usepackage{mciteplus}
\usepackage{placeins}
\newcommand{\eps}{\varepsilon}
\newcommand{\eV}{\ensuremath{\text{e}\mspace{-0.8mu}\text{V}\xspace}}

\newcommand{\GeV}{\ensuremath{\text{G\eV}}\xspace}
\newcommand{\TeV}{\ensuremath{\text{T\eV}}\xspace}

\newcommand{\pt}{\ensuremath{p_\perp}\xspace}

\newcommand{\mur}{\ensuremath{\mu_\text{R}}\xspace}
\newcommand{\muf}{\ensuremath{\mu_\text{F}}\xspace}

\newcommand{\sherpa}{\textsc{Sherpa}\xspace}

\newcommand{\bbbar}{\ensuremath{b\bar{b}}}

\makeatletter

\g@addto@macro\bfseries{\boldmath}
\makeatother

\begin{document}
\preprint{FERMILAB-PUB-19-148-T, MCNET-19-07}

\title{Multi-jet merging in a variable flavor number scheme}

\author{Stefan~H{\"o}che}
\affiliation{Fermi National Accelerator Laboratory,
  Batavia, IL, 60510-0500, USA}
\author{Johannes Krause}
\affiliation{Institut f{\"u}r Kern- und Teilchenphysik, TU Dresden,
  D-01062 Dresden, Germany}
\author{Frank Siegert}
\affiliation{Institut f{\"u}r Kern- und Teilchenphysik, TU Dresden,
  D-01062 Dresden, Germany}

\begin{abstract}
  We propose a novel technique for the combination of multi-jet merged
  simulations in the five-flavor scheme with calculations for the production
  of $b$-quark associated final states in the four-flavor scheme.
  We show the equivalence of our algorithm to the FONLL method at the
  fixed-order and logarithmic accuracy inherent to the matrix-element and
  parton-shower simulation employed in the multi-jet merging.
  As a first application we discuss $Zb\bar{b}$ production at the
  Large Hadron Collider.
\end{abstract}

\maketitle

\section{Introduction}

Measurements involving heavy-flavor (HF) production are a vital
component of the physics program at the Large Hadron Collider
(LHC). With the Higgs boson decaying predominantly into $b$ quarks,
some of the recent efforts in the LHC experiments have focused
on this decay mode in Higgs production in association with
vector bosons~\cite{Aaboud:2018zhk,Sirunyan:2018kst} and in
association with top quarks~\cite{Sirunyan:2018hoz,Aaboud:2017rss}.
Furthermore, searches for physics beyond the Standard Model (SM)
also often rely on heavy flavor final states because the couplings
to third generation fermions are sometimes assumed to be enhanced
in new physics models.
While the modeling of signal processes is relevant, it is of even
higher importance to have a precise prediction for the dominant SM
backgrounds including heavy flavor production. This is reflected
for example by the large efforts spent in the LHC Higgs Cross Section
Working Group~\cite{deFlorian:2016spz} to understand the modeling
of the $t\bar{t}b\bar{b}$ background to $t\bar{t}H(b\bar{b})$.

Including heavy-quark mass effects in the QCD evolution, e.g.\
in fits of parton distribution functions (PDFs), is a prerequisite.
The fixed flavor number scheme (FFNS) as the simplest approach assumes
a fixed number of active quarks which can be varied explicitly~\cite{Alekhin:2009ni}.
General-mass variable flavor number schemes (VFNS) like ACOT~\cite{Aivazis:1993kh,Aivazis:1993pi,Tung:2001mv,Kniehl:2011bk},
TR~\cite{Thorne:1997ga,Thorne:2006qt}, and FONLL~\cite{Cacciari:1998it,Forte:2010ta} on the other hand account
for mass effects dynamically above the corresponding thresholds. Hybrid schemes have also been devised~\cite{Kusina:2013slm},
and the possibility to perform the PDF evolution for massive quarks has been investigated recently~\cite{Krauss:2017wmx}.
Accordingly, higher-order corrections to the production of a final state like $Hb\bar{b}$ or $Zb\bar{b}$
have been computed for a fixed number of flavors~\cite{Dittmaier:2003ej,Dawson:2003kb,FebresCordero:2008ci,Cordero:2009kv,Caola:2011pz},
and in variable flavor number schemes~\cite{Campbell:2008hh,Forte:2015hba,Forte:2016sja,Forte:2018ovl,Bonvini:2015pxa,Bonvini:2016fgf}.
Progress in clarifying the interplay between different schemes was reviewed in~\cite{Cordero:2015sba},
and the effect of higher-order electroweak corrections was investigated recently~\cite{Figueroa:2018chn}.

For experimental analyses a realistic simulation of collision events
at the hadron level is crucial. Such simulations are provided by all
modern Monte Carlo event generators~\cite{Buckley:2011ms} like
Herwig7~\cite{Bellm:2015jjp}, Pythia8~\cite{Sjostrand:2014zea}
and Sherpa~\cite{Gleisberg:2008ta}. The simulation of heavy-flavor
production employed in these programs varies both in method and in
accuracy, and a formal comparison to the methods used in analytical
calculations is missing so far.
Typical event generator setups employ matrix elements in the
five-flavor scheme (5FS), where the $b$-quark is treated as massless
and included also as an initial state parton.\footnote{Note that formally
  this would require the usage of FFNS PDFs.} Before the parton shower
is simulated, $b$-quark masses are restored by means of a kinematics
reshuffling, and subsequently massive splitting kernels and kinematics
are used in the parton shower, see e.g.~\cite{Norrbin:2000uu,
  Schumann:2007mg,Hoche:2015sya,Cabouat:2017rzi,Cormier:2018tog}.
This procedure is necessary in particular to avoid an excess in the
$g\to bb$ splitting rate as it would appear if the fragmentation process
was simulated with massless $b$-quarks.

To increase the accuracy of heavy-flavor production Monte Carlo
samples, two approaches have been studied in the literature:
The matching of NLO QCD calculations in the four-flavor scheme (4FS)
to parton showers using one of the common NLO+PS matching
formalisms~\cite{Frixione:2002ik,Nason:2004rx,Frixione:2007vw,Hoeche:2011fd,Platzer:2011bc}.
This method has been used to simulate for example the $pp\to Vbb$ process
class~\cite{Frederix:2011qg,Oleari:2011ey,Bagnaschi:2018dnh,Krauss:2016orf}
and the $pp\to ttbb$
process~\cite{Cascioli:2013era,Jezo:2018yaf,Bevilacqua:2017cru}.
An alternative option to include higher-order QCD corrections in MCs
is the use of multi-leg merging methods at
LO~\cite{Catani:2001cc,Mangano:2001xp,Krauss:2002up,Lonnblad:2001iq,Lavesson:2007uu,Alwall:2007fs,Hoeche:2009rj,Hamilton:2009ne,Lonnblad:2011xx}
or
NLO~\cite{Hoeche:2012yf,Gehrmann:2012yg,Lonnblad:2012ix,Frederix:2012ps,Bellm:2017ktr}
accuracy. Heavy-flavor production is then included automatically in
the 5FS using massless matrix elements and a massive parton shower as
described above. Dedicated studies of this and a comparison to the 4FS
can be found in~\cite{Krauss:2016orf}.

In this paper we propose a novel method that presents a hybrid between
these two approaches. Embedding a massive NLO+PS calculation for
$pp\to Zbb$ into a massless multi-leg merging of $pp\to Z$+jets we
retain the theoretical advantages of both methods and for the first
time allow a rigorous combination of the two calculations without
overlap. The latter is not merely a technical or academic point, but
crucial to allow the usage of NLO-accurate heavy flavor predictions in
experiments: If heavy flavor and light jet production can not be described
simultaneously, it is impossible to make predictions in the presence of
fake heavy flavor jets or evaluate experimental efficiencies related
to them.

This paper is organized as follows: After a short review of multi-jet
merging in Sec.~\ref{sec:review}, we describe our method from an
algorithmic point of view in Sec.~\ref{sec:fusing}. Its formal
relation to the FONLL method is studied in
Sec.~\ref{sec:fonll}. Finally, in Sec.~\ref{sec:application} we
demonstrate an implementation of the new method within the Sherpa
event generator using $pp\to Zb\bar{b}$ production as a test case.

\section{Multi-jet merging in a fixed flavor number scheme}
\label{sec:review}
In the context of Monte-Carlo event generators, multi-jet merging
refers to algorithms that systematically improve the accuracy of
traditional \mbox{LO+PS} simulations by adding higher multiplicity
fixed-order calculations with well-separated parton-level jets to
the simulation. Merging algorithms are constructed such as to obtain a
consistent result that preserves both the logarithmic accuracy of
the parton shower and the fixed order accuracy of all higher-order
results. Such a treatment becomes important if kinematics and
correlations between jets have to be accurately predicted.  In the
\sherpa event generator, multi-jet merging is implemented at leading
and next-to-leading order QCD accuracy, using the
MEPS@LO~\cite{Hoeche:2009rj} and
MEPS@NLO~\cite{Hoeche:2012yf,Gehrmann:2012yg} method,
respectively. Both methods are explained in full detail in their
respective references; here we briefly summarize the main ideas that
are relevant to extend them to a variable number of parton flavors.

The combination of resummation and higher-order perturbative calculations by merging involves two aspects:
\begin{enumerate}
\item The phase space of resolvable emissions in the resummation must be restricted to the complement
  of the phase space of the fixed-order calculation. For example, in the combination of $pp\to Z$
  and $pp\to Zj$, with $p_{T,j}>p_{T,\rm cut}$, the phase space of the first emission in the resummation
  would be restricted to $\pt<p_{\perp,\rm cut}$. This restriction is called the {\it jet veto},
  the variable used to separate the phase space is called the {\it jet criterion},\footnote{
    The jet criterion can be thought of as a jet resolution scale. It is constructed such as to identify
    configurations in which the matrix elements develop soft or collinear singularities.}
  and the separation scale is called the {\it merging scale}.
\item The fixed-order result must be amended by the resummed higher-order corrections in order to maintain
  the logarithmic accuracy in the overall calculation. This is formally relevant only if the fixed-order
  calculation is used in a region of phase space where resummation is both relevant and reliable,
  i.e.\ for merging scales smaller than the resummation scale. This procedure consists of
  \begin{enumerate}
  \item Re-interpreting the final-state configuration of the fixed-order calculation
    as having originated from a parton cascade~\cite{Andre:1997vh}. This procedure is called
    {\it clustering}, and the representations of the final-state configuration in terms of
    parton branchings are called {\it parton-shower histories}.
  \item Choosing appropriate scales for evaluating the strong coupling in each branching
    of this cascade, thereby resumming higher-order corrections to soft-gluon
    radiation~\cite{Amati:1980ch,Catani:1990rr}.%
    \footnote{We will refer to this scale definition as the MEPS scale in the following.}
    This procedure is called {\it $\alpha_s$-reweighting}.
  \item Multiplying by appropriate Sudakov factors, representing the resummed unresolved
    real and virtual corrections~\cite{Catani:2001cc}. This procedure is called
    {\it Sudakov reweighting}, and it is usually implemented using
    {\it pseudo-showers}~\cite{Lonnblad:2001iq}.
  \end{enumerate}
\end{enumerate}
The jet clustering procedure terminates when either no more combination of particles according to the
QCD Feynman rules can be performed, or when the scale hierarchy would be violated by a new combination.
Among all possible parton-shower histories that can be constructed, one is chosen probabilistically
according to the associated weight, to represent the event. This weight is computed as the product of
the weight of the irreducible {\rm core process} left after the clustering has terminated, multiplied
by the differential radiation probability at each of the nodes of the branching tree. Note that these
probabilities depend on the parton-shower algorithm. 

\begin{figure}[t]
  \includegraphics[scale=0.333]{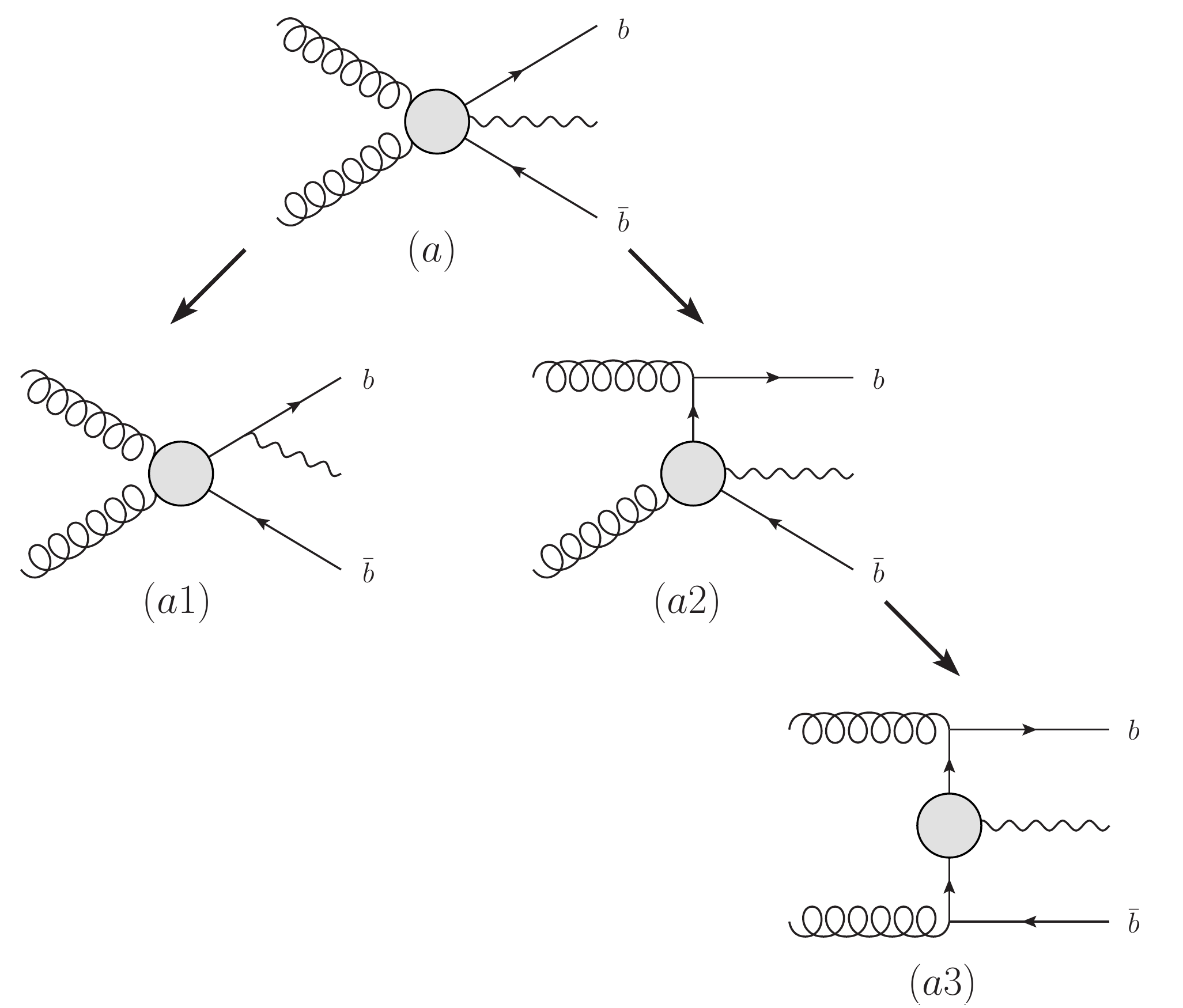}\hfill
  \includegraphics[scale=0.333]{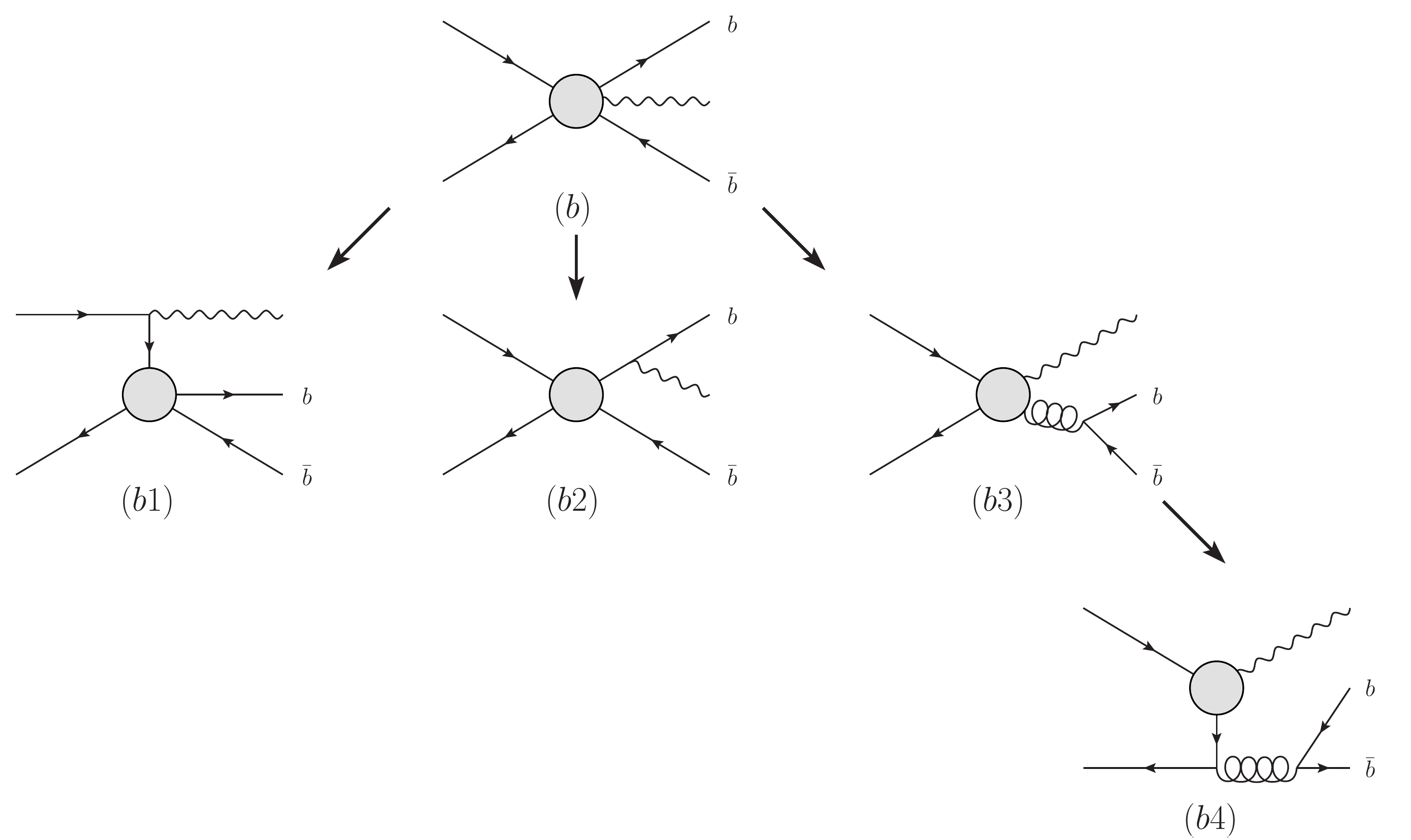}
  \caption{Representative parton-shower histories for $gg\to Zb\bar{b}$ (a) and $q\bar{q}\to Zb\bar{b}$ (b)
    matrix-element configurations. The gray blobs correspond to the irreducible core processes.}
  \label{fig:clustering}
\end{figure}
Some representative parton-shower histories, together with different core processes, are shown in
Fig.~\ref{fig:clustering}. Figure~\ref{fig:clustering}~(a) is the starting point of the jet clustering
procedure in $gg\to Zb\bar{b}$ configurations, which correspond to a leading-order prediction for
$pp\to Zb\bar{b}$ in the four-flavor scheme. The first QCD clustering that can be performed on this
configuration is the combination of a final-state (anti-)quark and an initial-state gluon, as indicated
in Fig.~\ref{fig:clustering}~(a2). The second would then be the combination of the other final-state
(anti-)quark and the second initial-state gluon, leading to Fig.~\ref{fig:clustering}~(a3).
The core process in this case is $b\bar{b}\to Z$, which corresponds to the lowest-order hard
matrix element in the five-flavor scheme. This example shows how the construction of parton-shower
histories from hard matrix elements provides a natural matching of the two schemes. We will expand
on this idea in Secs.~\ref{sec:fusing} and~\ref{sec:fonll}. If the scale in the second clustering
step, leading from Fig.~\ref{fig:clustering}~(a2) to Fig.~\ref{fig:clustering}~(a3) is lower
than the scale in the first step, we speak of a violation of the scale hierarchy in the merging.
In this case, the clustering procedure terminates after the first branching, and the core process
is $gb\to Zb$ ($g\bar{b}\to Z\bar{b}$). It is also possible that the scale in the first clustering
step exceeds all subsequently defined scales, including the scale associated with the core process.
In this case, it may still be possible to perform an electroweak clustering, leading to
Fig.~\ref{fig:clustering}~(a1), and defining the core process $gg\to b\bar{b}$. The correct
probability for this would be given by electroweak evolution equations~\cite{Bauer:2017isx,
  Bauer:2018arx,Fornal:2018znf}, which are not yet implemented in standard parton showers.
Therefore, one can choose to either neglect this clustering path (``exclusive clustering''),
or allow it using ad-hoc clustering probabilities (``inclusive clustering'').
Eventually, if no clustering can be performed
due to the scale hierarchy, the core process may correspond to the starting configuration
in Fig.~\ref{fig:clustering}~(a).
Similar arguments apply to the possible parton-shower histories shown in Fig.~\ref{fig:clustering}~right.
Again, the four-flavor scheme expression would correspond to Fig.~\ref{fig:clustering}~(b),
and the analogue in the five-flavor scheme would be Fig.~\ref{fig:clustering}~(b4).

\section{Multi-jet merging in a variable flavor number scheme}
\label{sec:fusing}

In this section we describe our new algorithm, which combines a merged calculation
in the five-flavor scheme with a prediction for heavy quark associated production.
Both the merged and the heavy flavor prediction may be computed at leading order
or at next-to-leading order QCD.
The combination is achieved by means of a dedicated heavy flavor overlap removal.
It acts on top of multi-jet merging algorithms and we call this technique {\it fusing}.
We first explain it from a phenomenological point of view, using the example of $Z$+jets /
$Zb\bar{b}$ production. The formal connection to the FONLL method will be established
in Sec.~\ref{sec:fonll}.

The basic idea of the fusing approach is as follows:
\begin{enumerate}
\item Start with a merged simulation of the inclusive reaction, e.g.\ $Z$+jets
  and a calculation of heavy quark associated production, e.g.\ $Zb\bar{b}$.
\item\label{alg:fusing_2}
  Process the $Zb\bar{b}$ simulation as if it was part of the multi-jet
  merged computation, i.e.\ apply the clustering, the $\alpha_s$ reweighting
  and the Sudakov reweighting.
  The renormalization and factorization scales for the core process should be
  calculated using a custom scale definition, and the scales of all reconstructed
  splittings should be set to the transverse momenta in the branching~\cite{Amati:1980ch},
  including higher-order corrections to soft-gluon evolution~\cite{Catani:1990rr},
  cf.\ Sec.~\ref{sec:review}.
  This part of the fused result will be called the {\it direct} component,
  as the final-state bottom quarks are generated in the fixed-order calculation.
\item\label{alg:fusing_3}
  Remove all final-state configurations from the five-flavor scheme merged
  simulation of $Z$+jets that have a parton-shower history
  which can also be generated in the reweighted $Zb\bar{b}$ computation.
  The remainder of the five-flavor scheme result may still contribute
  configurations with final-state bottom quarks. This part of the fused result will be called
  the {\it fragmentation} component.\footnote{The bottom quarks may be produced both
    in the fixed-order and in the parton-shower component of the merged result.
    As such, the expression {\it fragmentation contribution} is a slight misnomer.
    It is a true fragmentation contribution if the maximum jet multiplicity
    in the $Z$+jets calculation does not exceed the final-state multiplicity
    in the $Zb\bar{b}$ associated calculation.}
\item Add the modified event samples to obtain the fused result.
\end{enumerate}
\begin{figure}[t]
  \includegraphics[scale=0.333]{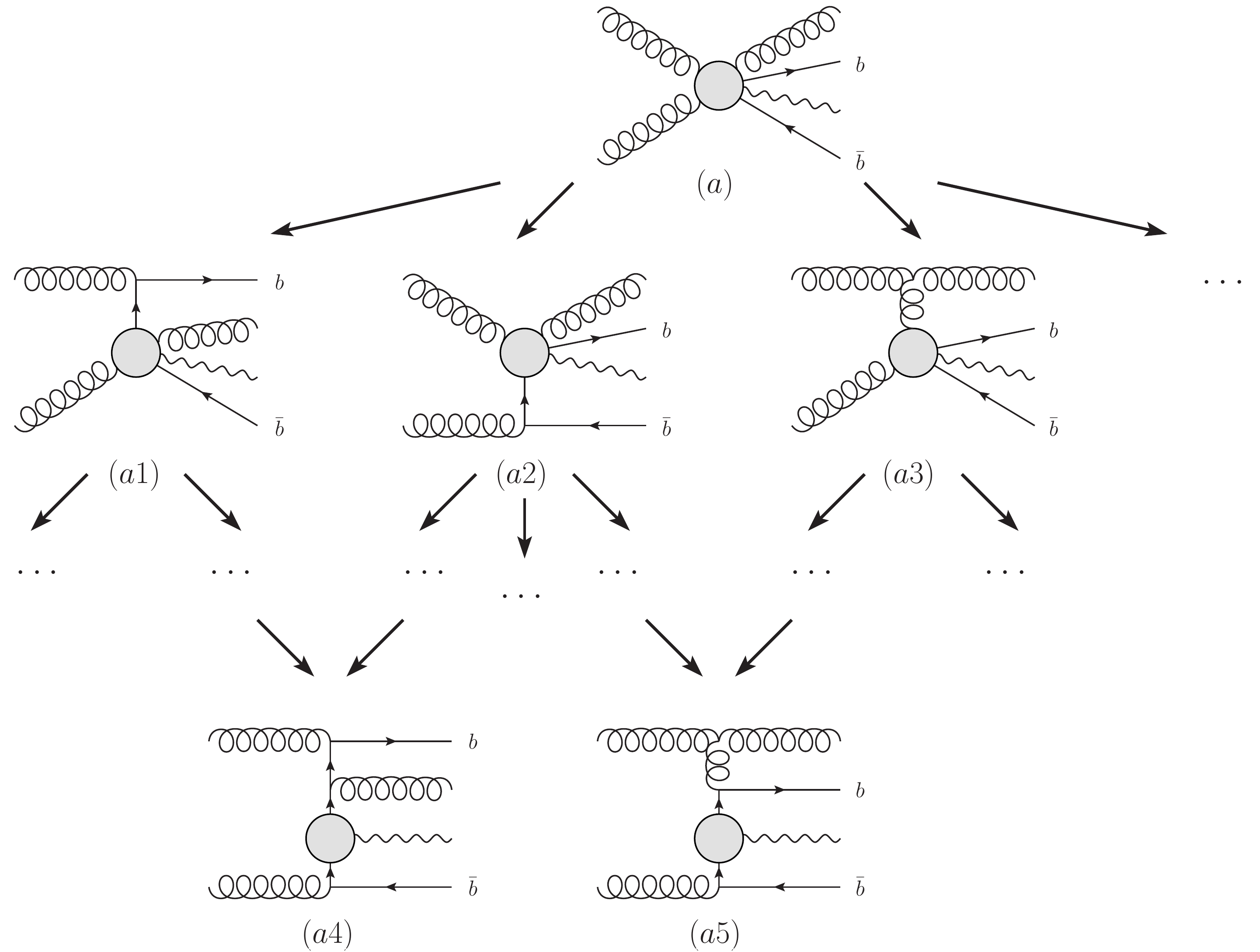}\hfill
  \includegraphics[scale=0.333]{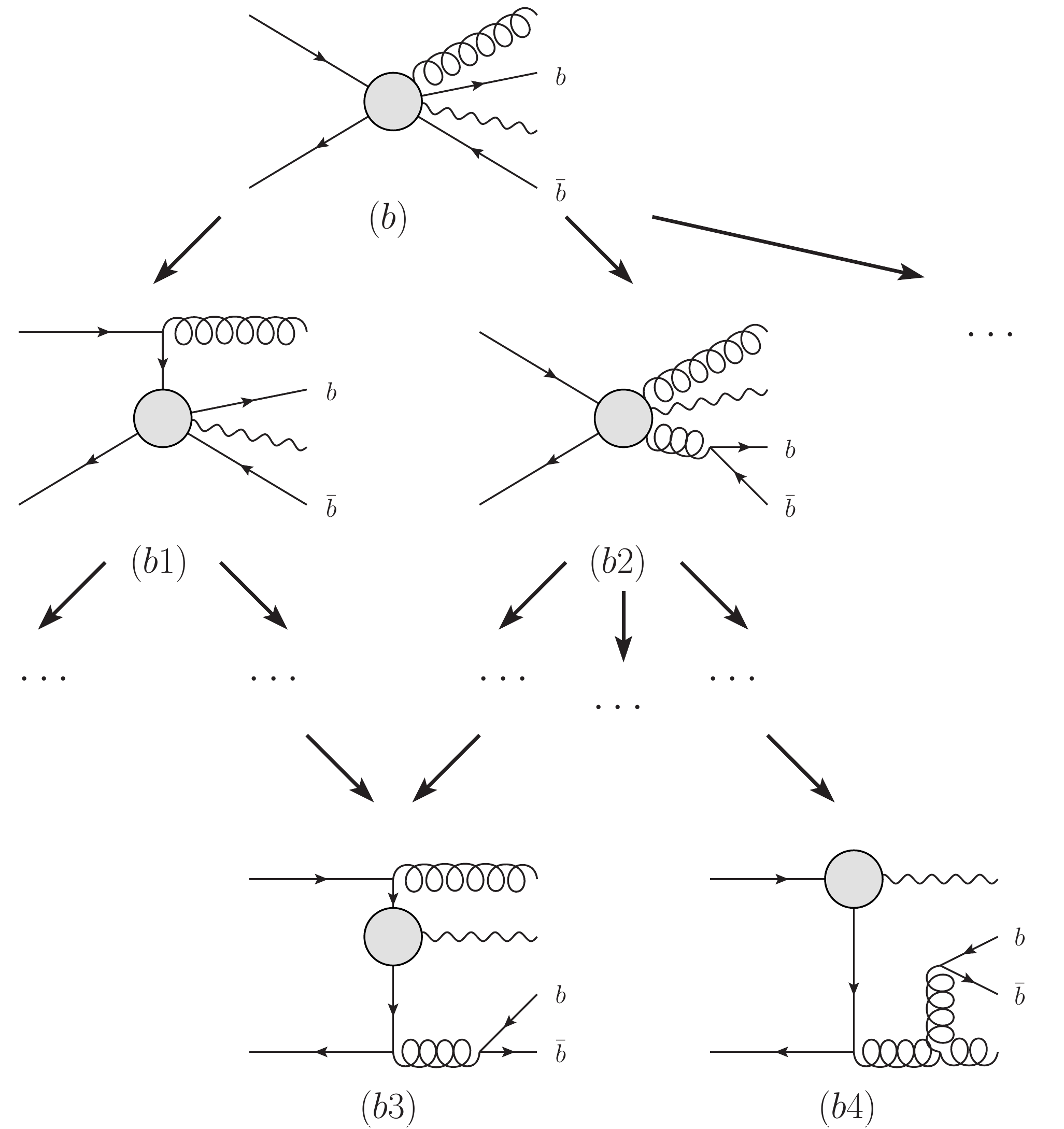}
  \caption{Example parton-shower histories for $gg\to Zb\bar{b}g$ (a),
    and $q\bar{q}\to Zb\bar{b}g$ (b). Depending on the clustering path,
    configurations~(a5) and (b3) may be identified with a $Zb\bar{b}$
    topology at leading order, while configurations~(a4) and (b4) may not.
    At next-to-leading order, all configurations can be identified with a
    $Zb\bar{b}$ topology.}
  \label{fig:fusing}
\end{figure}
The removal of overlap between the $Z$+jets and the $Zb\bar{b}$ calculations
is eventually achieved by both the Sudakov reweighting in step~\ref{alg:fusing_2}
and the event rejection in step~\ref{alg:fusing_3}.
The application of Sudakov vetoes to $Zb\bar{b}$ restores the correct behavior
of the direct component in those regions of phase space that feature a hierarchy
between the hard scale and the b-quark mass. The event rejection in the $Z$+jets
sample removes those final-state configurations which would otherwise be double
counted. Algorithmically this rejection is performed as follows:
\begin{enumerate}
\item Create a combined evolution history, starting from the core process
  in the jet clustering, and ending with all final state particles produced
  either in the hard matrix element or by the shower.
\item Starting from the core process, find the first configuration
  where a \bbbar-pair appears in the final-state.
\item If there is no such configuration, keep the event.
  Otherwise, count the number of additional light partons $n_{\text{light}}$
  (quarks or gluons) in the final state of this configuration.
  This corresponds to the number of hard emissions before the \bbbar-pair
  production according to the ordering imposed by the cluster (shower) algorithm.
  At leading order, discard the event if $n_{\text{light}}=0$.
  At next-to-leading order, discard the event if $n_{\text{light}}\le 1$.
\end{enumerate}
Typical parton-shower histories for candidate $Z$+jets events with gluon or 
quark initial states are shown in Fig.~\ref{fig:fusing}.
If the clustering leading to configurations (a5) / (b3)
proceeds along (a3) / (b1), the scale associated with the gluon emission is
the smallest in the process. The configuration can then be identified with
a $Zb\bar{b}$ topology, and the event will be discarded.
If the clustering proceeds along (a2) / (b2), the treatment depends on whether
the fusing is performed at leading or at next-to-leading order. At leading order
the configuration cannot be identified with a $Zb\bar{b}$ topology,
and the event will be kept. At next-to-leading order, the configuration
corresponds to a real-emission configuration, and the event will be discarded.
Figure~\ref{fig:fusing}~(a4) / (b4) displays a parton-shower history that does not have
a counterpart in the heavy-flavor result at leading order, such that the corresponding
event would be kept irrespective of the scale hierarchy. At next-to-leading order,
the event would be discarded.
The extension to histories with more partons in the final state is
straightforward and will lead to configurations that contribute to the
fragmentation component also in the next-to-leading order case.

Special care has to be taken when dealing with \emph{unordered} configurations. 
They arise when the clustering algorithm can not reconstruct
a strictly $k_T$-ordered history leading to a $pp\to Z$ core process
(cf.\ Sec.~\ref{sec:review}). In such cases, the clustering can either stop with a
$2\rightarrow n$ core process, or continue by allowing to violate the scale hierarchy.
Both variants can be used for the fusing algorithm as long as all components and all
parton multiplicities are treated identically. In this work we restrict ourselves
to a fully-ordered clustering algorithm.

\section{Relation to the FONLL method}
\label{sec:fonll}
This section will establish the relation between our merging algorithm and the
FONLL method~\cite{Forte:2015hba,Forte:2016sja}. In the FONLL technique,
the cross section of the combined event sample is generated as
\begin{equation}\label{eq:fonll_basic}
  \sigma^{\rm FONLL}=\sigma^{(5)}-\sigma^{(4),(0)}+\sigma^{(4)}\;,
\end{equation}
where $\sigma^{(5)}$ and $\sigma^{(4)}$ are the cross sections in the five- and four-flavor scheme,
respectively, and $\sigma^{(4),(0)}$ is the four-flavor scheme result in the limit $m_b\to0$.
Eventually, all results should only depend on the PDFs and strong coupling in the five-flavor scheme.
Formally, this is achieved by writing the cross section as
\begin{equation}
  \sigma^{(4)}=\int dx_1\int dx_2\,\sum_{ij=q,g}f_i^{(5)}(x_1,Q^2)f_j^{(5)}(x_2,Q^2)\,
  B_{ij}\left(x_1x_2,\alpha_s^{(5)}(Q^2),\frac{Q^2}{m_b^2}\right)\;.
\end{equation}
The hard coefficients $B_{ij}$ can then be expanded in powers of the strong coupling as
\begin{equation}
  B_{ij}\left(\tau,\alpha_s^{(5)}(Q^2),\frac{Q^2}{m_b^2}\right)
  =\sum_{n=2}\bigg(\frac{\alpha_s^{(5)}(Q^2)}{2\pi}\bigg)^n
  B_{ij}^{(n)}\left(\tau,\frac{Q^2}{m_b^2}\right)\;,
\end{equation}
and are determined such that the four-flavor scheme result in terms of four-flavor PDFs
is eventually recovered at the target accuracy given by the upper limit of the sum.

The coefficients $\sigma^{(4),(0)}$, needed for removal of the overlap
between the fully massive and the massless calculation, can be extracted from the
five-flavor scheme result~\cite{Forte:2016sja} by expressing the b-quark PDF up to $\mathcal{O}(\alpha_s)$
in terms of the four-flavor scheme light quark and gluon PDFs using the matching coefficients
from~\cite{Buza:1996wv}, and subsequently re-expressing the result in terms of five-flavor scheme
PDFs and $\alpha_s$~\cite{Forte:2010ta}. The result is
\begin{equation}\label{eq:fb}
  \begin{split}
  f_b^{(5)}(x,Q^2)=&\;\frac{\alpha_s(\mu_R^2)}{2\pi}
  \int_x^1\frac{dz}{z}A_{gb}^{(1)}(z,L)f_g^{(5)}\left(\frac{x}{z},Q^2\right)\\
  &\;+\frac{\alpha_s^2(\mu_R^2)}{(2\pi)^2}
  \int_x^1\frac{dz}{z}\left[\,A_{gb}^{(2)}(z,L)f_g^{(5)}\left(\frac{x}{z},Q^2\right)
  +A_{\Sigma b}^{(2)}(z,L)f_\Sigma^{(5)}\left(\frac{x}{z},Q^2\right)\,\right]\;,
  \end{split}
\end{equation}
where $L=\ln Q^2/m_b^2$ and $f_\Sigma=\sum_{a=q,\bar{q}}f_a(x,Q^2)$.
We can now use the $\mathcal{O}(\alpha_s^n)$ five-flavor scheme partonic cross sections
$\hat{\sigma}^{(n)}$ to define the massless limit of the coefficient functions $B^{(n)}$.
In the processes of interest to us the partonic cross section is invariant under exchange
of $b$ and $\bar{b}$. The $\mathcal{O}(\alpha_s^2)$ terms are then given by
\begin{equation}\label{eq:fonll_s402}
  \begin{split}
    B_{q\bar{q}}^{(0),(2)}\left(\tau,\frac{Q^2}{m_b^2}\right)=&\;\hat{\sigma}_{q\bar{q}}^{(2)}(\tau)
    +2\int_\tau^1\frac{dz}{z}A_{qq,b}^{(2)}(z,L)\,
    \hat{\sigma}_{q\bar{q}}^{(0)}\Big(\frac{\tau}{z}\Big)\;,\\
    B_{gg}^{(0),(2)}\left(\tau,\frac{Q^2}{m_b^2}\right)=&\;\hat{\sigma}_{gg}^{(2)}(\tau)
    +4\int_\tau^1\frac{dz}{z}A_{gb}^{(1)}(z,L)\,\hat{\sigma}_{gb}^{(1)}\left(\frac{\tau}{z}\right)
    +2\int_\tau^1\frac{dz}{z}\int_z^1\frac{dy}{y}A_{gb}^{(1)}(y,L)A_{gb}^{(1)}\Big(\frac{z}{y},L\Big)\,
    \hat{\sigma}_{b\bar{b}}^{(0)}\Big(\frac{\tau}{z}\Big)\;.
  \end{split}
\end{equation}
The $\mathcal{O}(\alpha_s^3)$ terms are given by
\begin{equation}\label{eq:fonll_s403}
  \begin{split}
    &B_{gg}^{(0),(3)}\left(\tau,\frac{Q^2}{m_b^2}\right)=
    \int_\tau^1\frac{dz}{z}\bigg[\,4A_{gb}^{(1)}(z,L)\,\hat{\sigma}_{gb}^{(2)}\left(\frac{\tau}{z},L\right)
      +4A_{gb}^{(2)}(z,L)\,\hat{\sigma}_{gb}^{(1)}\left(\frac{\tau}{z},L\right)\bigg]\\
    &\qquad+\int_\tau^1\frac{dz}{z}
    \int_z^1\frac{dy}{y}\bigg[\,2A_{gb}^{(1)}(y,L)A_{gb}^{(1)}\Big(\frac{z}{y},L\Big)\,
      \hat{\sigma}_{b\bar{b}}^{(1)}\left(\frac{\tau}{z}\right)
      +4A_{gb}^{(2)}(y,L)A_{gb}^{(1)}\Big(\frac{z}{y},L\Big)\,
      \hat{\sigma}_{b\bar{b}}^{(0)}\Big(\frac{\tau}{z}\Big)\bigg]\;,\\
    &B_{gq}^{(0),(3)}\left(\tau,\frac{Q^2}{m_b^2}\right)=\;
    \int_\tau^1\frac{dz}{z}\,
    \bigg[2A_{gb}^{(1)}(z,L)\,\hat{\sigma}_{qb}^{(2)}\left(\frac{\tau}{z},L\right)
      +2A_{\Sigma b}^{(2)}(z,L)\,\hat{\sigma}_{gb}^{(1)}\left(\frac{\tau}{z},L\right)\bigg]\\
    &\qquad+\int_\tau^1\frac{dz}{z}\int_{z}^1\frac{dy}{y}\,
    2A_{\Sigma b}^{(2)}(y,L)A_{gb}^{(1)}\Big(\frac{z}{y},L\Big)\,
    \hat{\sigma}_{b\bar{b}}^{(0)}\left(\frac{\tau}{z}\right)
    +\int_\tau^1\frac{dz}{z}\,2A_{qq,b}^{(2)}(z,L)\,\hat{\sigma}_{gq}^{(1)}\left(\frac{\tau}{z},L\right)\;,\\
    &B_{q\bar{q}}^{(0),(3)}\left(\tau,\frac{Q^2}{m_b^2}\right)=
    \int_\tau^1\frac{dz}{z}\,2A_{qq,b}^{(2)}(z,L)\,\hat{\sigma}_{q\bar{q}}^{(1)}\left(\frac{\tau}{z},L\right)\;.\\
  \end{split}
\end{equation}
The matching coefficients in Eq.~\eqref{eq:fb} can be expanded in a power series in $L$ as
\begin{equation}\label{eq:matching_coefficients}
  \begin{split}
    A_{gb}^{(1)}(z,L)=&\;a_{gb}^{(1,1)}(z)L\;,\\
    A_{gb}^{(2)}(z,L)=&\;a_{gb}^{(2,2)}(z)L^2+a_{gb}^{(2,1)}(z)L+a_{gb}^{(2,0)}(z)\;,\\
    A_{\Sigma b}^{(2)}(z,L)=&\;a_{\Sigma b}^{(2,2)}(z)L^2+a_{\Sigma b}^{(2,1)}(z)L+a_{\Sigma b}^{(2,0)}(z)\;,\\
    A_{qq,b}^{(2)}(z,L)=&\;a_{qq,b}^{(2,2)}(z)L^2+a_{qq,b}^{(2,1)}(z)L+a_{qq,b}^{(2,0)}(z)\;.
  \end{split}
\end{equation}
In the parton-shower approach, each logarithm $L$ arises from integrating
Eq.~\eqref{eq:pdf_evolution_constrained_2} over $\ln t$ from $m_b^2$ to $Q^2$.
By comparing to Eq.~\eqref{eq:fb} we find that $A_{gb}^{(1)}(z,L)=P_{gq}(z)L$.
We will comment on the remaining coefficients in Sec.~\ref{sec:fonll_nlonll}.
The non-logarithmic terms, $a^{(2,0)}(z)$, are needed only for matching beyond
NLL accuracy and can therefore be ignored in our approach. The leading and sub-leading
logarithmic terms can be derived from renormalization and collinear mass factorization
of the operator matrix elements for heavy quark production~\cite{Buza:1995ie}.
They take the simple form
\begin{equation}\label{eq:fonll_coefficients}
  \begin{split}
  a_{gb}^{(1,1)}(z)=&\;P_{gq}(z)\;,\\
  a_{gb}^{(2,2)}(z)=&\;\frac{1}{2}\int_z^1\frac{dx}{x}P_{gq}(x)P_{qq}\left(\frac{z}{x}\right)
    +\beta_0\,P_{gq}(z)-\frac{1}{2}\int_z^1\frac{dx}{x}
    P_{gg}(x)P_{gq}\left(\frac{z}{x}\right)\;,
  &a_{gb}^{(2,1)}(z)=&\;P_{gq}^{(1)}(x)\;,\\
  a_{\Sigma b}^{(2,2)}(z)=&\;-\frac{1}{2}\int_z^1\frac{dx}{x}P_{qg}(x)P_{gq}\left(\frac{z}{x}\right)\;,
  &a_{\Sigma b}^{(2,1)}(z)=&\;P_{qq}^{S,(1)}(x)\;,\\
  a_{qq,b}^{(2,2)}(z)=&\;-\frac{1}{2}\,\beta_{0,b}\,P_{qq}(z)\;,
  &a_{qq,b}^{(2,1)}(z)=&\;P_{qq,Q}^{V,(1)}(x)\;.
  \end{split}
\end{equation}
The leading-order and relevant next-to-leading order splitting functions
entering Eqs.~\eqref{eq:fonll_coefficients} are given in App.~\ref{sec:sfs}.
The negative term in $a_{gb}^{(2,2)}(z)$ and the coefficient $a_{\Sigma b}^{(2,2)}(z)$
are collinear mass factorization counterterms that arise from the different number
of quark flavors in the infrared and the ultraviolet regime~\cite{Buza:1995ie,Buza:1996wv}.

\subsection{Leading Order and Leading Logarithmic Accuracy}
\label{sec:fonll_loll}
In order to prove that the heavy flavor overlap removal algorithm proposed
in this publication amounts to a variant of the FONLL method we need to show
that the removal of events from the five-flavor sample as proposed in
Sec.~\ref{sec:fusing} is equivalent to the subtraction of $\sigma^{(4),(0)}$
in the FONLL technique. We will start at leading order and
leading logarithmic accuracy and comment on next-to-leading
order and next-to-leading logarithmic accuracy in the next section.

The simplest configurations are $\hat{\sigma}^{(2)}_{q\bar{q}}$ and $\hat{\sigma}^{(2)}_{gg}$
in Eq.~\eqref{eq:fonll_s402}. They correspond to removal of the double-real radiative
corrections to the $b\bar{b}\to Z$ process, which have a counterpart in the four-flavor scheme.
Note that, in the notation of Eq.~\eqref{eq:fonll_s402}, $\hat{\sigma}^{(2)}_{q\bar{q}}$ and
$\hat{\sigma}^{(2)}_{gg}$ are integrated over the double-real radiative phase space
and combined with the renormalized virtual corrections and collinear mass factorization
counterterms, which renders both $\hat{\sigma}^{(2)}_{q\bar{q}}$ and $\hat{\sigma}^{(2)}_{gg}$
individually finite.
In the $\overline{\rm MS}$ scheme, the factorization scale dependent remainder combines
with the PDF evolution to give the second and third expression on the right-hand side
of Eq.~\eqref{eq:fonll_s402}. In a multi-jet merging approach, no singularities arise
because we effectively use a cutoff regulator for collinear mass singularities that
is defined by the jet cuts. For $gg$ initial states we obtain
\begin{equation}\label{eq:meps_s402}
  \begin{split}
    B_{gg,\rm MEPS}^{(0),(2)}\left(\tau,\frac{Q^2}{m_b^2}\right)=&\;
    \int{\rm d}\Phi_{2}\,\frac{{\rm d}\hat{\sigma}_{gg}^{(2)}(\tau)}{{\rm d}\Phi_2}\,
    \Theta(Q_1-Q_{\rm cut})\Theta(Q_2-Q_{\rm cut})\\
    &+4\int_{m_b^2}^{Q^2}\!\!\frac{{\rm d}t}{t}\int_\tau^1\frac{dz}{z}P_{gq}(z)\,
    \int{\rm d}\Phi_{1}\,\frac{{\rm d}\hat{\sigma}_{gb}^{(1)}\left(\tau/z\right)}{{\rm d}\Phi_1}\,
    \Theta(Q_1-Q_{\rm cut})\Theta(Q_{\rm cut}-Q_2)\\
  &+2\int_{m_b^2}^{Q^2}\!\!\frac{{\rm d}t}{t}\int_\tau^1\frac{dz}{z}
  \int_{m_b^2}^{t}\!\!\frac{{\rm d}t'}{t'}\int_z^1\frac{dy}{y}P_{gq}(y)P_{gq}\Big(\frac{z}{y}\Big)\,
  \hat{\sigma}_{b\bar{b}}^{(0)}\Big(\frac{\tau}{z}\Big)\,
  \Theta(Q_{\rm cut}-Q_1)\\
  \approx&\;B_{gg}^{(0),(2)}\left(\tau,\frac{Q^2}{m_b^2}\right)\bigg|_{\;\rm LL}\;,
  \end{split}
\end{equation}
where the subscript ${\rm LL}$ indicates leading logarithmic accuracy.
The scales $Q_1$ and $Q_2$ denote the jet resolution in the final and next-to-final
clustering of the merging algorithm, while $Q_{\rm cut}$ stands for the merging scale.
The $\Theta$-functions represent the phase-space partitioning in the merging procedure
with at least two jets in addition to the production of the inclusive final state
in the five-flavor scheme. The last approximation is valid if the merging cut is small
enough that below it we can factorize $\hat{\sigma}_{gg}^{(2)}$ and $\hat{\sigma}_{gb}^{(1)}$
into $\hat{\sigma}_{b\bar{b}}^{(0)}$ and $P_{gb}$, and if we can ignore the finite
remainder of the virtual corrections included in Eq.~\eqref{eq:fonll_s402}.
The subtraction of $\sigma^{(4),(0)}$ from $\sigma^{(5)}$ in Eq.~\eqref{eq:fonll_basic}
can therefore be achieved by using the algorithm in Sec.~\ref{sec:fusing}.
In the case of $gg$ initial states it proceeds as follows:
\begin{enumerate}
\item Construct a parton-shower history according to the multi-jet merging procedure,
  perform the parton shower and add any splittings that were generated to the history
\item\label{loll_algo_2} Starting at the core interaction identified in the merging,
  trace the parton-shower history. Veto the event if
  \begin{enumerate}
  \item the core process is $b\bar{b}\to Z$, followed by
    an initial-state $g\to b\bar{b}$ and $g\to \bar{b}b$ branching
  \item the core process is $gb\to Zb$ ($g\bar{b}\to Z\bar{b}$), followed
    by an initial-state $g\to b\bar{b}$ ($g\to \bar{b}b$) branching
  \item the core process is $gg\to Zb\bar{b}$
  \end{enumerate}
\end{enumerate}
The solution is similar for quark initial states, only the sequence
of parton-shower splittings differs. The multi-jet merged expression reads
\begin{equation}\label{eq:meps_s402_2}
  \begin{split}
    B_{q\bar{q},\rm MEPS}^{(0),(2)}\left(\tau,\frac{Q^2}{m_b^2}\right)=&\;
    \int{\rm d}\Phi_{2}\,\frac{{\rm d}\hat{\sigma}_{q\bar{q}}^{(2)}(\tau)}{{\rm d}\Phi_2}\,
    \Theta(Q_2-Q_{\rm cut})\Theta(Q_1-Q_{\rm cut})\\
    &+\int_{m_b^2}^{Q^2}\!\!\frac{{\rm d}t}{t}\int_0^1 {\rm d}z\,z\,P_{gq}(z)\,
    \int{\rm d}\Phi_{1}\,\frac{{\rm d}\hat{\sigma}_{q\bar{q}}^{(1)}\left(\tau\right)}{{\rm d}\Phi_1}\,
    \Theta(Q_1-Q_{\rm cut})\Theta(Q_{\rm cut}-Q_2)\\
    &+2\int_{m_b^2}^{Q^2}\!\!\frac{{\rm d}t}{t}\int_\tau^1\frac{dz}{z}
    \int_{m_b^2}^{t}\!\!\frac{{\rm d}t'}{t'}\int_0^1{\rm d}y\,y\,P_{gq}(y)P_{qq}(z)\,
    \hat{\sigma}_{b\bar{b}}^{(0)}\Big(\frac{\tau}{z}\Big)\,
    \Theta(Q_{\rm cut}-Q_1)\\
    \approx&\;B_{q\bar{q}}^{(0),(2)}\left(\tau,\frac{Q^2}{m_b^2}\right)\bigg|_{\;\rm LL}\;.
  \end{split}
\end{equation}
The origin of $a_{qq,b}^{(2,2)}(z)$ contained in
$B_{q\bar{q}}^{(0),(2)}\big|_{\;\rm LL}$
is explained in Sec.~\ref{sec:fonll_nlonll}.

\subsection{Next-to-Leading Order and Next-to-Leading Logarithmic Accuracy}
\label{sec:fonll_nlonll}
In order to achieve next-to-leading logarithmic accuracy according to the
FONLL method, the parton shower employed in the merging must implement all
coefficient functions in Eq.~\eqref{eq:fonll_coefficients}. We start with
the $\beta_0$ dependent contribution to $a_{gb}^{(2,2)}(z)$. Making use
of the expansion of the strong coupling in the four-flavor scheme to
$\mathcal{O}(\alpha_s)$,
\begin{equation}\label{eq:fonll_ps_a20}
  \alpha_s(m_b^2)=\alpha_s\left[\,1+\frac{\alpha_s}{2\pi}\,\beta_0\,L\,\right]\;,
\end{equation}
this term can either be implemented explicitly, or absorbed into the scale choice
connected to the evaluation of $a_{gb}^{(1,1)}(z)$. In the latter case,
the strong coupling in initial-state $g\to b\bar{b}$ splittings should be
computed at $m_b^2$. Because we use a strong coupling in the five-flavor scheme,
an additional counterterm of the form $\alpha_s/(2\pi)\beta_{0,b}L$ will then be
required.

The coefficients $a^{(2,2)}(z)$ can be expressed in the parton-shower
formalism as the convolution of $a_{gb}^{(1,1)}(z)$ with the emission and
no-emission probability of the parton shower, expanded to second order in the
strong coupling and integrated over the evolution parameter from $m_b^2$ to $Q^2$.
To show this, we employ the correspondence between inclusive and exclusive
parton evolution summarized in App.~\ref{app:nlo_vs_ps}.
Making use of Eq.~\eqref{eq:pdf_evolution_2} and the boundary condition
$f_b(x,m_b^2)=0$, a single step in the parton-shower backward evolution,
generating a resolved $g\to b\bar{b}$ transition at scale $Q^2$ can be
written formally as
\begin{equation}\label{eq:ps_proof_lo}
  f_b(x,Q^2)=\int_{m_b^2}^{Q^2}\!\frac{{\rm d}t}{t}\,\frac{\alpha_s}{2\pi}
  \int_x^1\frac{{\rm d} z}{z}\,P_{gq}(z)\,f_{g}\Big(\frac{x}{z},t\Big)\;.
\end{equation}
Upon expansion to $\mathcal{O}(\alpha_s)$, we obtain the leading-order
coefficient of Eq.~\eqref{eq:fb}. To reconstruct the first term in
$a_{gb}^{(2,2)}(z)$, we need to account for a second step in the
parton-shower evolution, preceding the $g\to b\bar{b}$ transition.
This gives
\begin{equation}
  \begin{split}
  &\int_{m_b^2}^{Q^2}\!\frac{{\rm d}t}{t}
  \,\frac{\alpha_s}{2\pi}\int_x^{1-\eps}\frac{{\rm d}z}{z}\left[
  P_{gq}(z)\,\Delta_q(t,Q^2)
  +\int_{t}^{Q^2}\!\frac{{\rm d}\bar{t}}{\bar{t}}\,\frac{\alpha_s}{2\pi}
  \int_z^{1-\eps'}\frac{{\rm d}y}{y}
  P_{gq}\Big(\frac{z}{y}\Big)\,\hat{P}_{qq}(y)\,
  \Delta_q(\bar{t},Q^2)\right]\,f_g\Big(\frac{x}{z},t\Big)\;.
  \end{split}
\end{equation}
Subtracting the leading-order term in Eq.~\eqref{eq:ps_proof_lo},
and expanding to second order in the strong coupling, we can write
\begin{equation}\label{eq:ps_fonll_a21_pre}
  \frac{\alpha_s^2}{(2\pi)^2}\int_{m_b^2}^{Q^2}\!\frac{{\rm d}t}{t}
  \int_x^{1-\eps}\frac{{\rm d}z}{z}\int_{t}^{Q^2}\frac{{\rm d}\bar{t}}{\bar{t}}\left[
    -P_{gq}(z)\sum_{a=q,g}\int_0^{1-\eps'}\!\!{\rm d}\zeta\,\zeta\,\hat{P}_{qa}(\zeta)
    +\int_z^{1-\eps'}\frac{{\rm d}y}{y}\,P_{gq}\Big(\frac{z}{y}\Big)\,\hat{P}_{qq}(y)\right]
  \,f_g\Big(\frac{x}{z},Q^2\Big)\;.
\end{equation}
Using Eq.~\eqref{eq:sf_regularization} and~\eqref{eq:kernels_fineps}, and taking the limit $\eps,\eps'\to 0$, gives
\begin{equation}
  \frac{\alpha_s^2}{(2\pi)^2}\int_{m_b^2}^{Q^2}\!\frac{{\rm d}t}{t}
  \int_x^1\frac{{\rm d}z}{z}\int_{t}^{Q^2}\frac{{\rm d}\bar{t}}{\bar{t}}
  \int_z^1\frac{{\rm d}y}{y}\,P_{gq}\Big(\frac{z}{y}\Big)\,P_{qq}(y)
  \,f_g\Big(\frac{x}{z},Q^2\Big)\;.
\end{equation}
Finally, we make use of the fact that at $\mathcal{O}(\alpha_s^2)$ there is
no further dependence on $t$ and $\bar{t}$. Integrating them out
we obtain the contribution of the first term in $a_{gb}^{(2,2)}(z)$
to Eq.~\eqref{eq:fb}
\begin{equation}\label{eq:ps_fonll_a21}
  \frac{\alpha_s^2}{(2\pi)^2}\frac{L^2}{2}
  \int_x^1\frac{{\rm d}z}{z}
  \int_z^1\frac{{\rm d}y}{y}\,P_{gq}\Big(\frac{z}{y}\Big)\,P_{qq}(y)
  \,f_g\Big(\frac{x}{z},Q^2\Big)\;.
\end{equation}
The coefficient $a_{\Sigma b}^{(2,2)}(z)$ and the final term in
$a_{gb}^{(2,2)}(z)$ are derived in a similar way. The difference
compared to the previous case is that the hierarchy between $m_b^2$ and
$Q^2$ is ill-defined, because the second branching happens at smaller scales
than the $g\to b\bar{b}$ transition. The complete parton-shower expression
for the splitting kernel in $a_{gb}^{(1,1)}(z)$ and one step of the subsequent
evolution reads
\begin{equation}
  \begin{split}
  &\int_{m_b^2}^{Q^2}\!\frac{{\rm d}t}{t}
  \,\frac{\alpha_s}{2\pi}\int_x^{1-\eps}\frac{{\rm d}z}{z}\left[
    P_{gq}(z)\,
    \Delta_g(q^2,t)
    +\sum_{a=q,g}\int_{q^2}^{t}\frac{{\rm d}\bar{t}}{\bar{t}}\frac{\alpha_s}{2\pi}
    \int_z^{1-\eps'}\frac{{\rm d}y}{y}
    \hat{P}_{ag}\Big(\frac{z}{y}\Big)\,P_{gq}(y)\,
    \Delta_g(\bar{t},t)
    \right]\,f_g\Big(\frac{x}{z},t\Big)\;.
  \end{split}
\end{equation}
Note that we have introduced an auxiliary scale, $q^2$, in order to perform
the integral over the second branching. Subtracting the leading-order term
in Eq.~\eqref{eq:ps_proof_lo}, and expanding to second order in the strong coupling,
we can write
\begin{equation}\label{eq:ps_fonll_a22_pre}
  \begin{split}
  \frac{\alpha_s^2}{(2\pi)^2}\int_{m_b^2}^{Q^2}\!\frac{{\rm d}t}{t}
  \int_x^{1-\eps}\frac{{\rm d}z}{z}\int^{t}_{q^2}\frac{{\rm d}\bar{t}}{\bar{t}}\Bigg[
    &-P_{gq}(z)\sum_{a=q,g}\int_0^{1-\eps'}\!\!{\rm d}\zeta\,\zeta\,\hat{P}_{ga}(\zeta)
    \,f_g\Big(\frac{x}{z},Q^2\Big)\\
    &+\sum_{a=q,g}\int_z^{1-\eps'}\frac{{\rm d}y}{y}\,\hat{P}_{ag}\Big(\frac{z}{y}\Big)\,P_{gq}(y)
    \,f_a\Big(\frac{x}{z},Q^2\Big)\Bigg]\;.
  \end{split}
\end{equation}
Using again Eq.~\eqref{eq:sf_regularization} and~\eqref{eq:kernels_fineps}, taking the limit $\eps,\eps'\to 0$,
and integrating over $t$ and $\bar{t}$, we obtain
\begin{equation}\label{eq:ps_fonll_a22_pre2}
  \frac{\alpha_s^2}{(2\pi)^2}\,\frac{L}{2}\ln\frac{m_b^2Q^2}{q^4}
  \int_x^1\frac{{\rm d}z}{z}\sum_{a=q,g}\int_z^1\frac{{\rm d}y}{y}\,
  P_{ag}\Big(\frac{z}{y}\Big)\,P_{gq}(y)
  \,f_a\Big(\frac{x}{z},Q^2\Big)\;.
\end{equation}
Note that $q^2$ plays the role of the collinear mass factorization scale,
while $Q^2$ corresponds to the UV renormalization scale. In the parton-shower
approach, the two are strictly ordered, as the second branching cannot take place
before the first one. In a fixed-order computation, we can instead set $q=Q$.
This corresponds to treating the UV renormalization and collinear mass factorization
counterterms in Eqs.~(3.15) and~(3.20) of~\cite{Buza:1995ie} on the same footing,
which is eventually mandated by the choice to set $\epsilon_{\rm UV}=\epsilon_{\rm IR}$.
Using the same scheme in Eq.~\eqref{eq:ps_fonll_a22_pre2}, i.e.\ setting $q\to Q$,
we obtain the contribution of $a_{\Sigma b}^{(2,2)}(z)$ and the final term in
$a_{gb}^{(2,2)}(z)$ to Eq.~\eqref{eq:fb}
\begin{equation}\label{eq:ps_fonll_a22}
  -\frac{\alpha_s^2}{(2\pi)^2}\,\frac{L^2}{2}
  \int_x^1\frac{{\rm d}z}{z}\sum_{a=q,g}\int_z^1\frac{{\rm d}y}{y}\,
  P_{ag}\Big(\frac{z}{y}\Big)\,P_{gq}(y)
  \,f_a\Big(\frac{x}{z},Q^2\Big)\;.
\end{equation}
The coefficient $a_{qq,b}^{(2,2)}(z)$ can be derived in a similar fashion.
The difference compared to $a_{\Sigma b}^{(2,2)}(z)$ lies in the fact that
the phase space for the final-state branching of the intermediate gluon
into $b\bar{b}$ can be integrated out, leading to the coefficient $-\beta_{0,b}$.
The complete parton-shower expression reads
\begin{equation}
  \begin{split}
  &\int_{m_b^2}^{Q^2}\!\frac{{\rm d}t}{t}
    \,\frac{\alpha_s}{2\pi}\sum_{a=b,\bar{b}}
    \int_0^{1-\eps'}{\rm d}y\,y\,P_{ga}(y)
    \Bigg[\Delta_q(t,Q^2)+\int^{Q^2}_{t}\frac{{\rm d}\bar{t}}{\bar{t}}
    \frac{\alpha_s}{2\pi}\int_x^{1-\eps}\frac{{\rm d}z}{z}P_{qq}(z)
    \Delta_q(\bar{t},Q^2)\Bigg]\,f_g\Big(\frac{x}{z},t\Big)\;.
  \end{split}
\end{equation}
This can be expanded to second order in the strong coupling as in
Eqs.~\eqref{eq:ps_fonll_a21_pre} and~\eqref{eq:ps_fonll_a22_pre}.
Taking the limit $\eps,\eps'\to 0$, and integrating over $t$ and $\bar{t}$,
we obtain the contribution of $a_{qq,b}^{(2,2)}(z)$ to Eq.~\eqref{eq:fb}
\begin{equation}\label{eq:ps_fonll_a23}
  \frac{\alpha_s^2}{(2\pi)^2}\,\frac{L^2}{2}\,
  \bigg(\sum_{a=b,\bar{b}}\int_0^1{\rm d}y\,y\,P_{ga}(y)\bigg)
  \int_x^1\frac{{\rm d}z}{z}\,P_{qq}(z)\,
  f_g\Big(\frac{x}{z},t\Big)=
  -\frac{\alpha_s^2}{(2\pi)^2}\,\frac{L^2}{2}\,\beta_{0,b}
  \int_x^1\frac{{\rm d}z}{z}\,P_{qq}(z)\,
  f_g\Big(\frac{x}{z},t\Big)\;.
\end{equation}
Combining the parton-shower effects leading to Eqs.~\eqref{eq:ps_fonll_a21},
\eqref{eq:ps_fonll_a22}, \eqref{eq:ps_fonll_a23} and~\eqref{eq:fonll_ps_a20},
we obtain all double logarithmic coefficients in Eq.~\eqref{eq:fonll_coefficients}.
The single logarithmic coefficients are not reproduced by a standard parton shower
and must be implemented separately.
In the case of $a_{\Sigma b}^{(2,1)}$ and $a_{qq,b}^{(2,1)}$ this can be achieved
using the algorithm derived in~\cite{Hoche:2017iem}. Although a complete Monte-Carlo
implementation has not yet been presented for $a_{gb}^{(2,1)}$, it is clear that it
can be constructed using the techniques of~\cite{Hoche:2017iem} and~\cite{Dulat:2018vuy}.
In the foreseeable future it will therefore be possible to achieve next-to-leading
logarithmic accuracy according to the FONLL classification by using our approach.

In order to achieve next-to-leading order accuracy according to the FONLL method,
we must reconstruct the coefficient functions in Eq.~\eqref{eq:fonll_s403}.
The modification of the leading-order result, Eq.~\eqref{eq:fonll_s402}, by the
multi-jet merging procedure has already been discussed in Sec.~\ref{sec:fonll_loll}.
In complete analogy, the terms proportional to $\hat{\sigma}_{b\bar{b}}$ in
Eq.~\eqref{eq:fonll_s403} are modified by $\Theta(Q_{\rm cut}-Q_1)\Theta(Q_{\rm cut}-Q_2)$
in the merging, while the terms proportional to $\hat{\sigma}_{gb}$ and $\hat{\sigma}_{qb}$
are modified by $\Theta(Q_1-Q_{\rm cut})\Theta(Q_{\rm cut}-Q_2)$.
It remains to adjust the result for the fact that the four-flavor scheme coefficients
in the perturbative expansion are determined using a strong coupling in the
five-flavor scheme. We use Eqs.~(8) and (9) of~\cite{Forte:2016sja} to correct
this mismatch. Technically this is achieved by modifying the event weight $w$
of four-flavor scheme $S$-events with $gg$ or $q\bar{q}$ initial-state as
\begin{equation}\label{eq:fonll_alphas}
  \begin{split}
    w_{q\bar{q}}^{\text{new}} =& w_{q\bar{q}}\left( 1 - \frac{4}{3}T_R\ln\frac{\mu_R^2}{Q^2} \frac{w^{\text{Born}}}{w^{\text{ME}}}\right)\\
    w_{gg}^{\text{new}} =& w_{gg}\left( 1 - \frac{4}{3}T_R\ln\frac{\mu_R^2}{m_b^2} \frac{w^{\text{Born}}}{w^{\text{ME}}} \right)\;,\\
  \end{split}
\end{equation}
where $w^{\text{Born}}$ and $w^{\text{ME}}$ are the matrix-element weights of
the Born contribution and the full $S$-event, respectively.

\section{Results}
\label{sec:application}

The algorithm described in Sec.~\ref{sec:fusing} has been implemented in the 
Sherpa event generator in full generality, and will be investigated in the 
following for the example of heavy flavor production in association with a $Z$ 
boson.

As argued in Section~\ref{sec:fonll_nlonll}, a NLO-accurate parton shower
would be needed to fully match the next-to-leading logarithmic accuracy of the FONLL method.
Such a shower is not available yet but recent studies indicate that the central predictions
provided by it should be close to the LO result~\cite{Hoche:2017iem,Dulat:2018vuy}.
We thus base our new approach on the established MEPS@NLO algorithm, which we have extended to 
provide the necessary fully-ordered clustering and combined parton-shower history
for Sherpa version~2.2.7.
The event filter for the overlap removal described in Sec.~\ref{sec:fusing}
and the counter-terms described in Eq.~\eqref{eq:fonll_alphas} are implemented as user hooks.
The event filter can either directly be used to veto events or store the veto information as alternative event
weight. This allows to produce a $Z$+jets sample which is usable both standalone,
or within a fused prediction by applying the corresponding event weight and
adding a dedicated sample only for the direct component.

In the next sections we compare predictions obtained by the newly developed algorithm against existing predictions in
the 5FS and the 4FS. In all of them the matrix elements are generated using Sherpa's internal 
matrix element generators \textsc{Amegic++}~\cite{Krauss:2001iv} and \textsc{Comix}~\cite{Gleisberg:2008fv}
for tree-level diagrams.
Virtual diagrams are interfaced from \mbox{\textsc{OpenLoops 1.3.1}}~\cite{Cascioli:2011va}, 
using  \mbox{\textsc{CutTools}}~\cite{Ossola:2007ax} and \mbox{\textsc{OneLoop}}~\cite{vanHameren:2010cp}.

In the 5FS prediction and for the fragmentation component, matrix elements are calculated for 
\mbox{$pp \rightarrow \ell^+ \ell^- + 0,1,2j\text{@NLO} + 3j\text{@LO}$}. Here, $\ell^+,\ell^-$ refers to either electrons or muons
and $j$ to a well separated parton. The merging cut $Q_\text{cut}$ is set to $20~\GeV$ if not stated otherwise.
The direct component and the standalone 4FS prediction are based on $ pp\rightarrow \ell^+ \ell^- b \bar{b}$ matrix elements at
next-to-leading order, matched to the parton shower using the formalism in~\cite{Hoeche:2011fd}.

For both components of the fusing approach, and for the standalone 5FS and 4FS predictions,
all scales are evaluated according to the METS scheme with inclusive clustering\footnote{Ad-hoc electroweak cluster
steps have found to be relevant if the $\pt$ of the $Z$-boson becomes of order $100~\GeV$.}. 
In all predictions except for the standalone 4FS prediction the clustering is required to be fully ordered.
The factorization and (where applicable) renormalization scales in the core process are evaluated according to
\begin{equation}
\label{eq:def_core_scale}
   \mu^2_\text{core}=\begin{cases}
       m^2_{\ell\ell}            & \text{for $Z$}, \\
    \tfrac{1}{4}  m^2_{\perp,\ell\ell}            & \text{for $Zj$},\\
    \tfrac{1}{4}\tfrac{-1}{1/\hat{s}+1/\hat{t}+1/\hat{u}}  & \text{for jet+jet},\\
    \tfrac{1}{4}\bigl(m_{\perp,\ell\ell} ~ + \sum_{\text{jets}} m_{\perp,\text{jet}} \bigr)^2  & \text{for unordered $Z$+jets.}\\
  \end{cases}
\end{equation}
We use the NNPDF3.0 set~\cite{Ball:2014uwa} at NNLO with five active flavors for both components of the fusing approach and for the 5FS prediction,
and with four active flavors for the 4FS prediction.
These PDF sets are interfaced to Sherpa using LHAPDF~\cite{Buckley:2014ana}.
We use the CS-Shower~\cite{Schumann:2007mg} as implemented in Sherpa with two minor modifications.
Firstly, the strong coupling in $g\to b\bar{b}$ splittings is evaluated at the virtuality of the
intermediate parton, to account for the fact that there is no soft gluon emission, and therefore
no higher-order corrections enhanced by $\alpha_s/(2\pi)\beta_0\ln k_T^2/Q^2$. Secondly, we choose
the evolution variables of Scheme~1 in~\cite{Hoeche:2014lxa}, but we add the squared masses of the
final-state partons in the branching.
The default multiple interactions~\cite{Sjostrand:1987su,*Alekhin:2005dx} and hadronization
models~\cite{Winter:2003tt} implemented in Sherpa are employed in all simulations. 
Analyses of the event samples are performed within the \textsc{Rivet} framework~\cite{Buckley:2010ar}.
Scale variations ($\muf$, $\mur$) are studied using the on-the-fly-variation method as implemented in Sherpa~\cite{Bothmann:2016nao}.

\subsection{Validation in inclusive $Z$ phase space}
\begin{figure}[t]
  \centering
  \subfloat[The transverse momentum of the $Z$ boson.]{\includegraphics[width=0.45\textwidth]{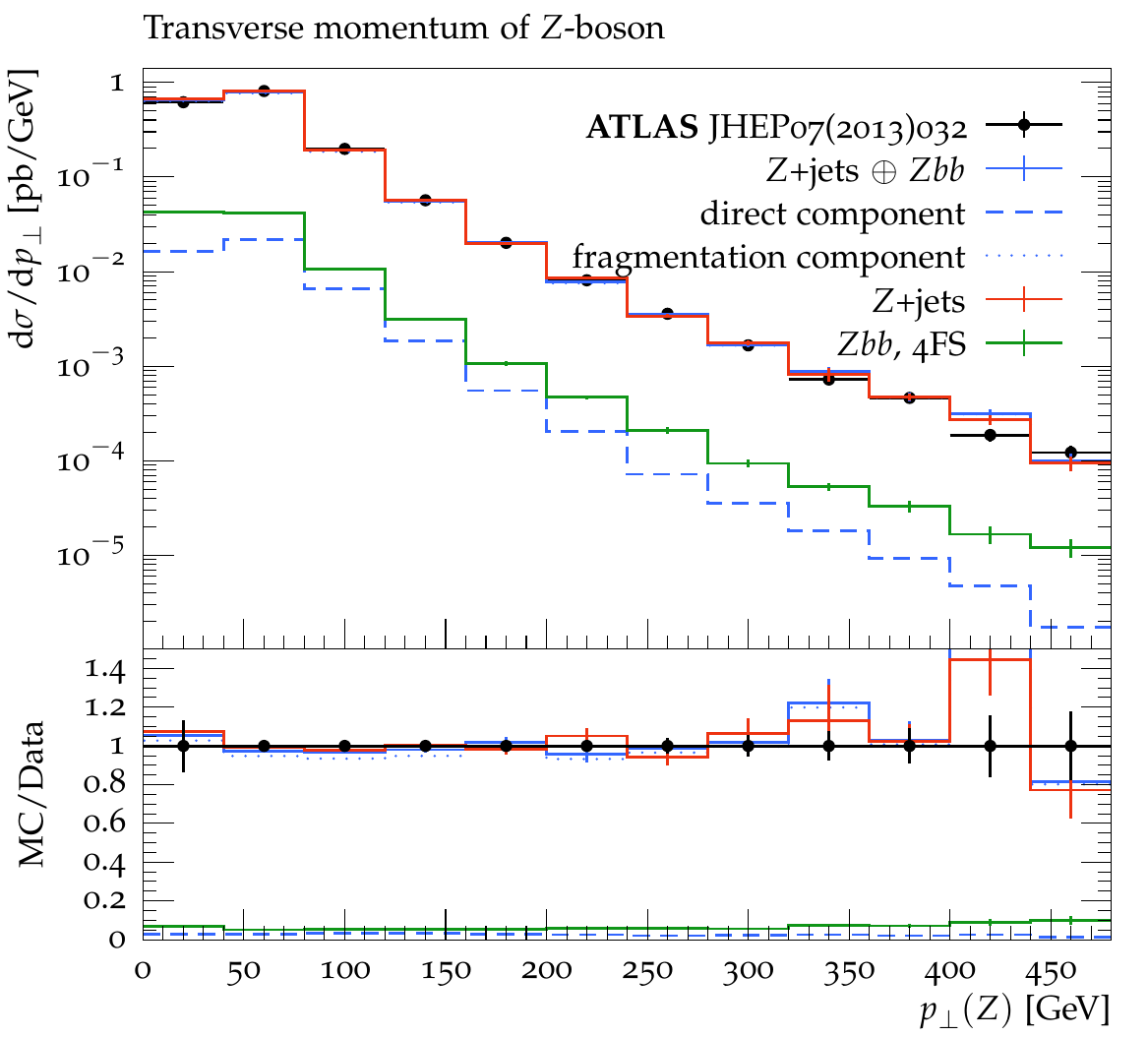}}
  \subfloat[The scalar sum of the transverse jet momenta, $S_T=\sum_i p_{\perp,i}^\text{jet}$.
	]{\includegraphics[width=0.45\textwidth]{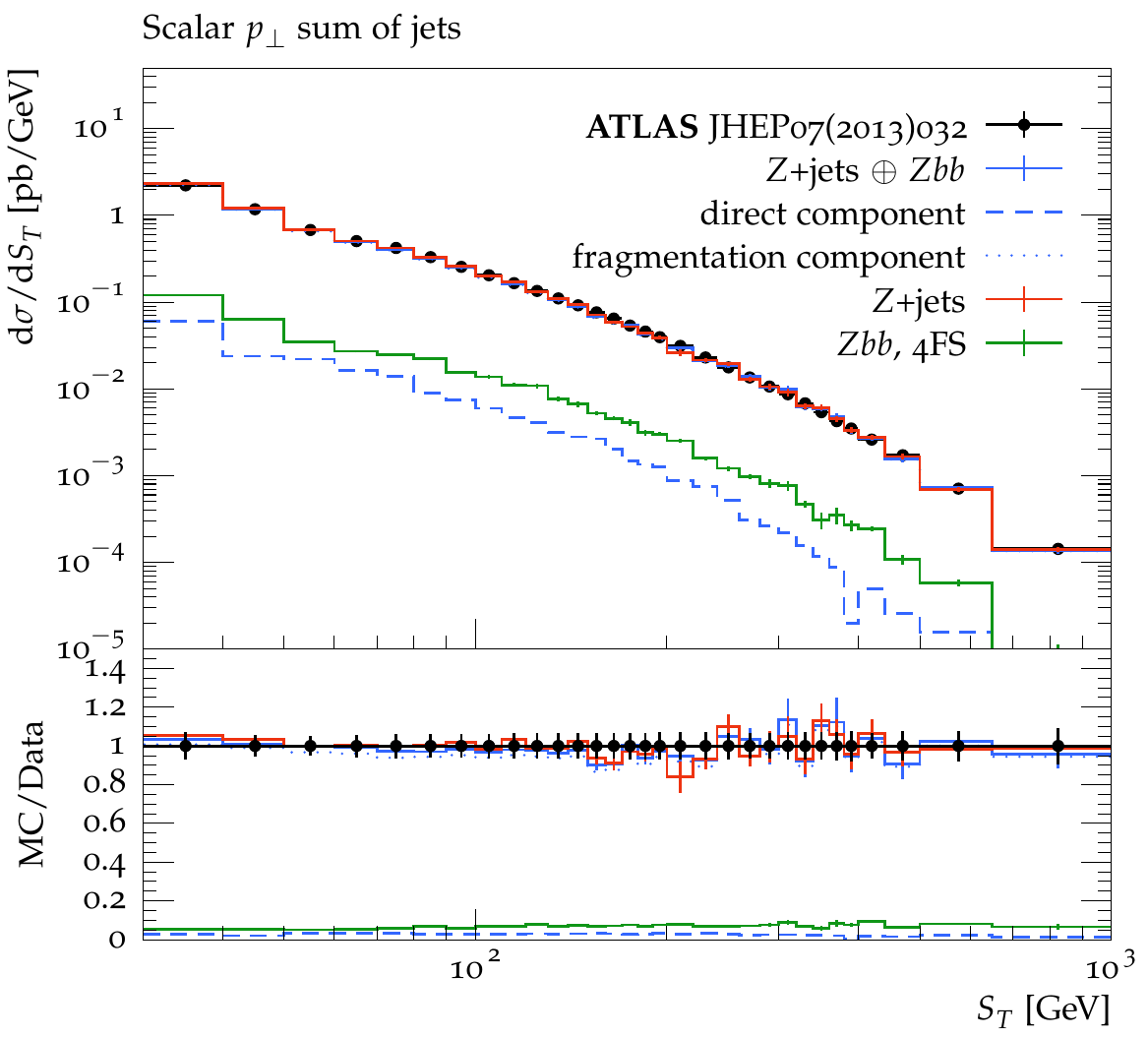}}
  \caption{Comparison of the newly developed algorithm with the established MEPS@NLO method and data from ATLAS in 
	    a $Z+\text{jets}$ region of phase space. The fused prediction is shown in solid blue, its both components
	    are given as dashed and dotted blue lines. The 5FS MEPS@NLO and the 4FS MC@NLO are shown for comparison in red
	    and green.} 
  \label{fig:z_inclusive}
\end{figure}

If the event selection does not explicitly require any $b$-jet,
our fusing approach is expected to agree with the MEPS@NLO prediction.
This is validated here with 7~\TeV data from ATLAS~\cite{Aad:2013ysa}.
In this measurement, $Z$ bosons decaying to electrons or muons were measured in association with jets.
Leptons are required to have a transverse momentum of $\pt^\ell>20~\GeV$ and a combined invariant mass with $66~\GeV < m_{\ell\ell} < 116~\GeV$. Jets are defined by the anti-$k_t$ algorithm with $R=0.4$ with a transverse momentum of $\pt^j>30~\GeV$ and a minimal angular distance to the leptons, 
$\Delta R(j,\ell)>0.5$.

%evaluation
In Fig.~\ref{fig:z_inclusive}, the transverse momentum spectrum $\pt^Z$ of the $Z$ boson and the scalar sum of
the jet transverse momenta, $S_T$, are shown.  
As expected, the fusing prediction is dominated by the fragmentation component in this region of phase space.
The full 4FS prediction still reaches up to 10~\% of the cross section, showing the necessity for a rigorous combination.
In both distributions, the new prediction is compatible
with the experimental data, and the agreement with the MEPS@NLO prediction demonstrates that the
fusing algorithm does not induce any unexpected features.
%  S_T, Z-pT

%\FloatBarrier

\subsection{Results in $Zbb$ phase space}

\begin{table}[t]
	\begin{tabularx}{\textwidth}{c|p{4cm} p{5cm} p{4cm} p{4cm} }
		\hline
									& \textbf{Data} [pb] & \textbf{Fusing} [pb] & \textbf{$Zbb$, 4F}  [pb] & \textbf{$Z+\text{jets}$} [pb] \\ 
		\hline
		\textbf{$Z + \ge 1 b$} & $3.55 \pm 0.24_\text{comb}$   & $3.80 \pm 0.05_\text{stat} \pm \begin{array}{l}   0.83 \\  0.33 \end{array} \tiny{\text{pert}} $  & $3.14 \pm 0.03_\text{stat}$  & $4.77 \pm 0.10_\text{stat}$ \\
		\textbf{$Z + \ge 2 b$} & $0.331 \pm 0.037_\text{comb}$ & $0.282 \pm 0.004_\text{stat} \pm \begin{array}{l}   0.027 \\  0.022 \end{array} \tiny{\text{pert}}$  & $0.305 \pm 0.006_\text{stat}$  & $0.358 \pm 0.012_\text{stat}$ \\
		\hline
	\end{tabularx}
        \caption{The total cross section for having at least one or at least two $b$-jets. Different predictions are compared to data from CMS.
			The uncertainties given for the data are combined uncertainties, the predictions are given with their statistical (all)
			and	perturbative (only fusing) uncertainties. }
	\label{tab:z_b_xs}
\end{table}

%1b-bin:  bjet pT,          z-pT,         HT,              dphi
%	d01-x01-y01     d05-x01-y01    d07-x01-y01      d09-x01-y01
\begin{figure}
  \centering
  \subfloat[The transverse momentum of the leading $b$-jet.]{\includegraphics[width=0.45\textwidth]{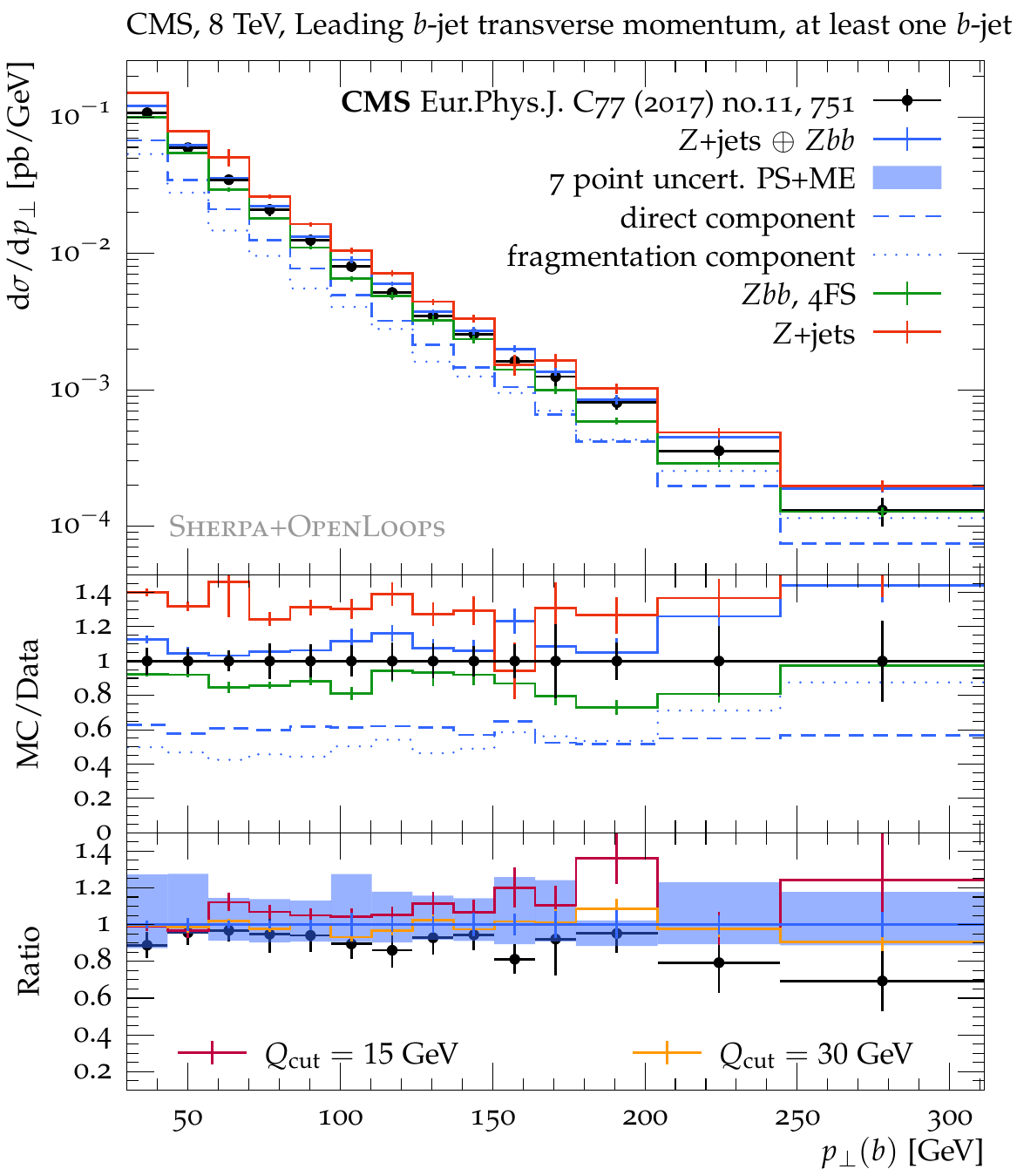}}
  \subfloat[The transverse momentum of the $Z$ boson.]{\includegraphics[width=0.45\textwidth]{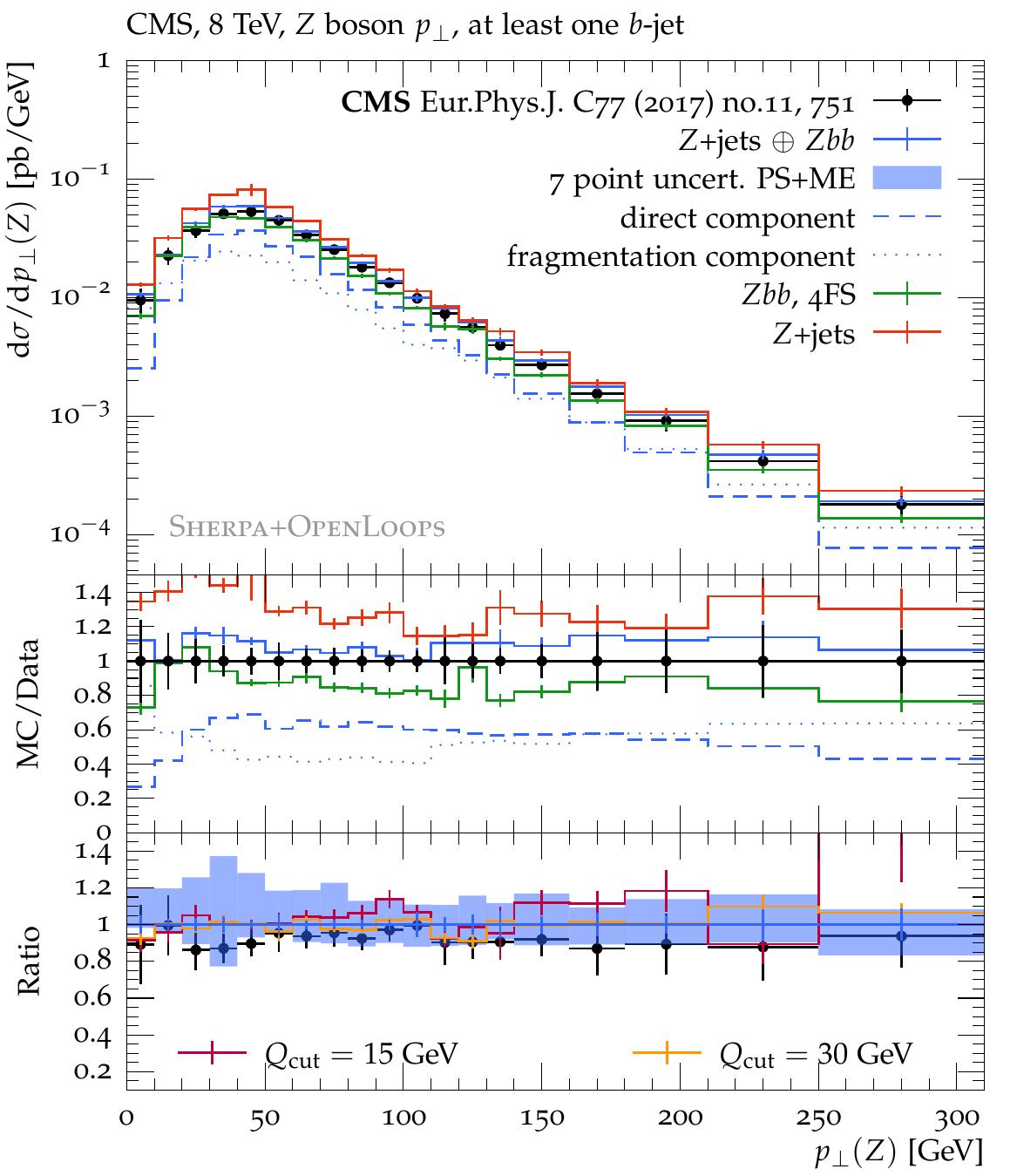}} \\
  \subfloat[The scalar sum of all transverse jet momenta, $S_T=\sum_i p_{\perp,i}^\text{jet}$. ]{\includegraphics[width=0.45\textwidth]{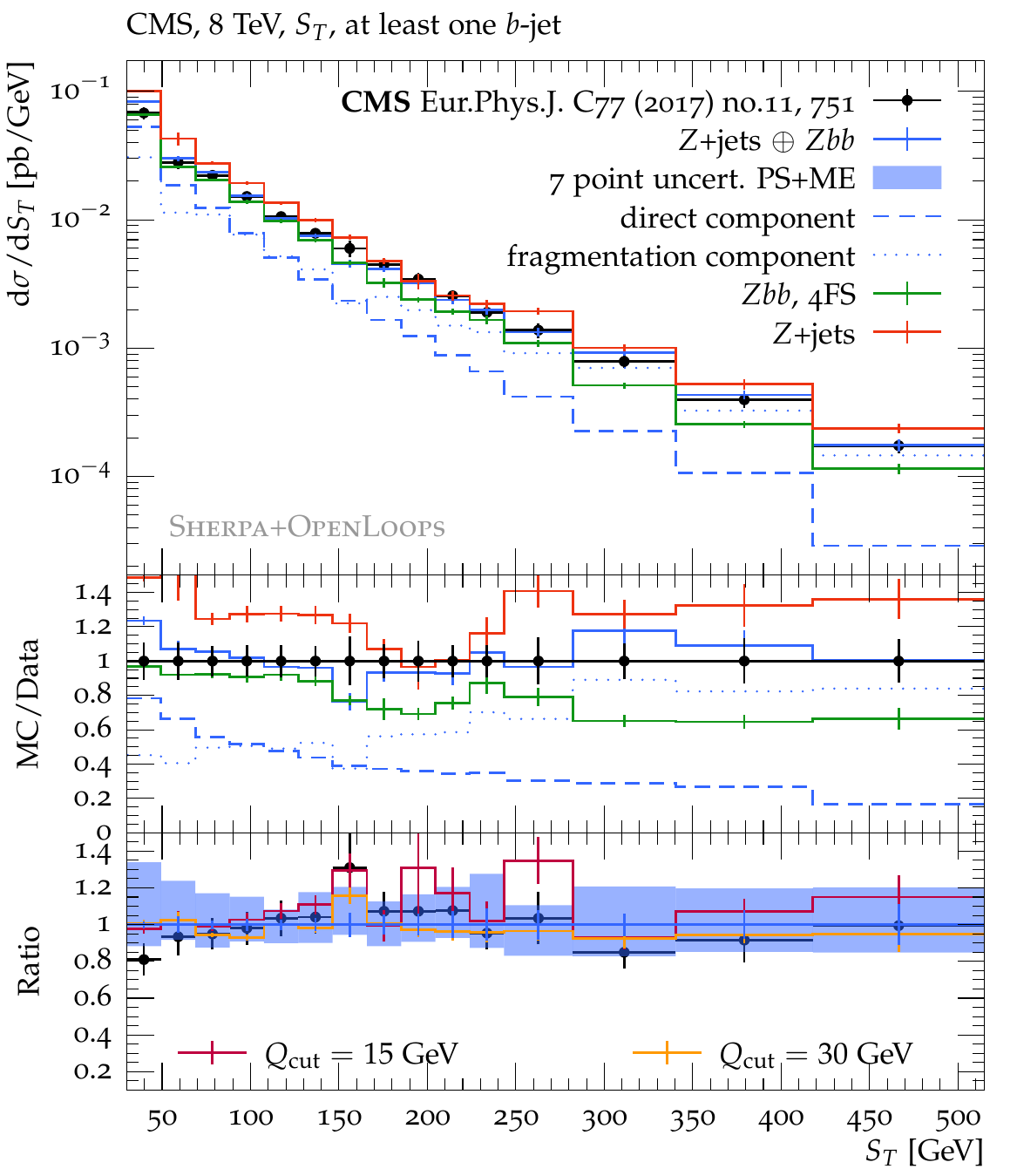}}
  \subfloat[The azimuthal distance between the leading $b$-jet and the $Z$-boson. ]{\includegraphics[width=0.45\textwidth]{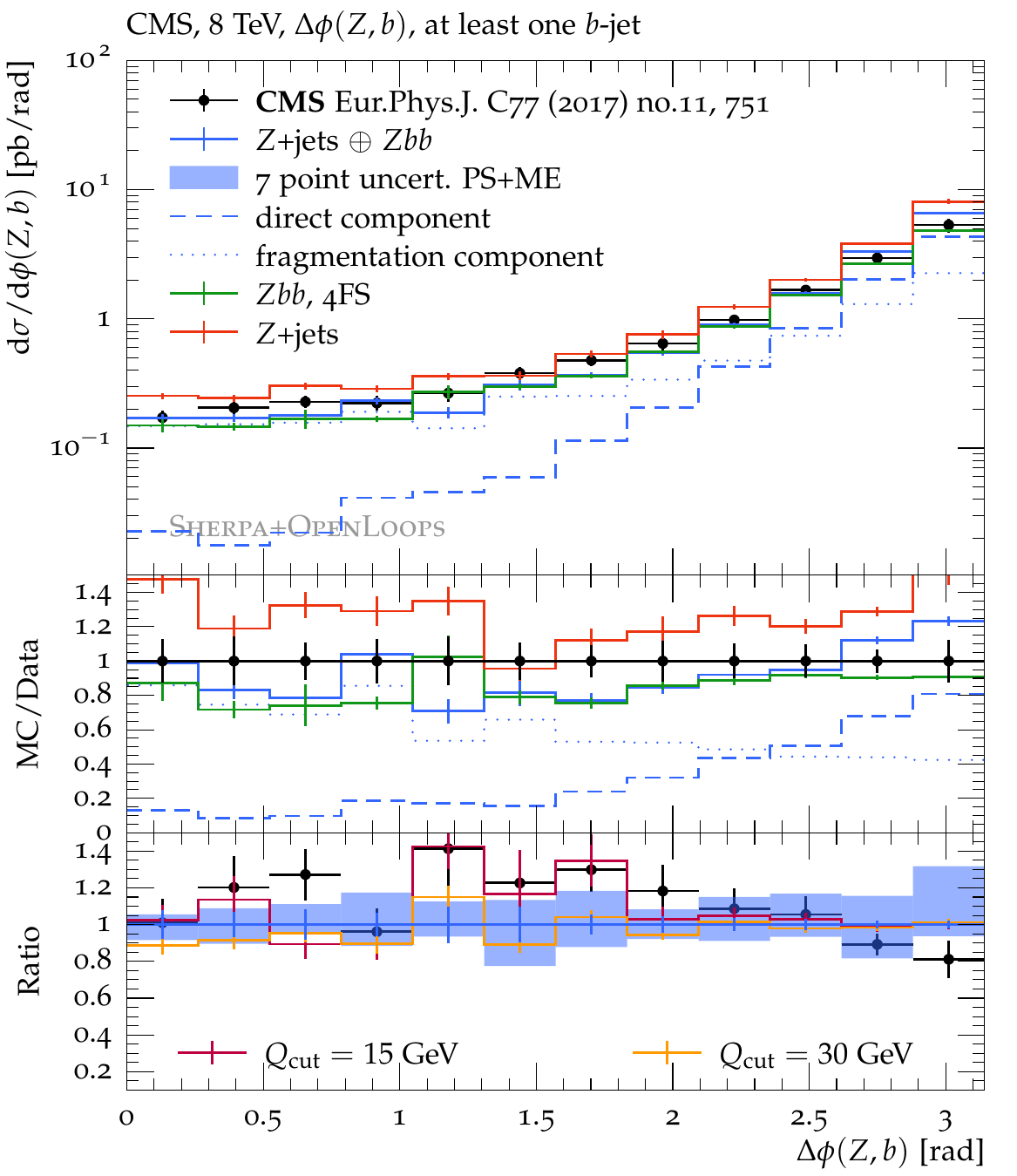}}
  \caption{Comparison of the newly developed algorithm with data from CMS in a phase space region with at least one $b$-jet. 
	    The new fusing prediction is shown in solid blue, its both components
	    are given as dashed and dotted blue lines. The 5FS MEPS@NLO and the 4FS MC@NLO are shown for comparison in red
	    and green. The uncertainties for the fused prediction are shown in the second ratio plot. 
	    They include a simultaneous seven-point variation of $\muf$ and $\mur$ for both, matrix element and parton shower emissions,
	    and a merging cut variation for $Q_\text{cut}=15(30)~\GeV$.} 
  \label{fig:z_one_b}
\end{figure}

%2b-bin:    M_bb,        dR_bb,       z pT            A_zbb
%       d14-x01-y01    d17-x01-y01    d13-x01-y01   d19-x01-y01
\begin{figure}
  \subfloat[The invariant mass distribution of the both leading $b$-jets.]{\includegraphics[width=0.45\textwidth]{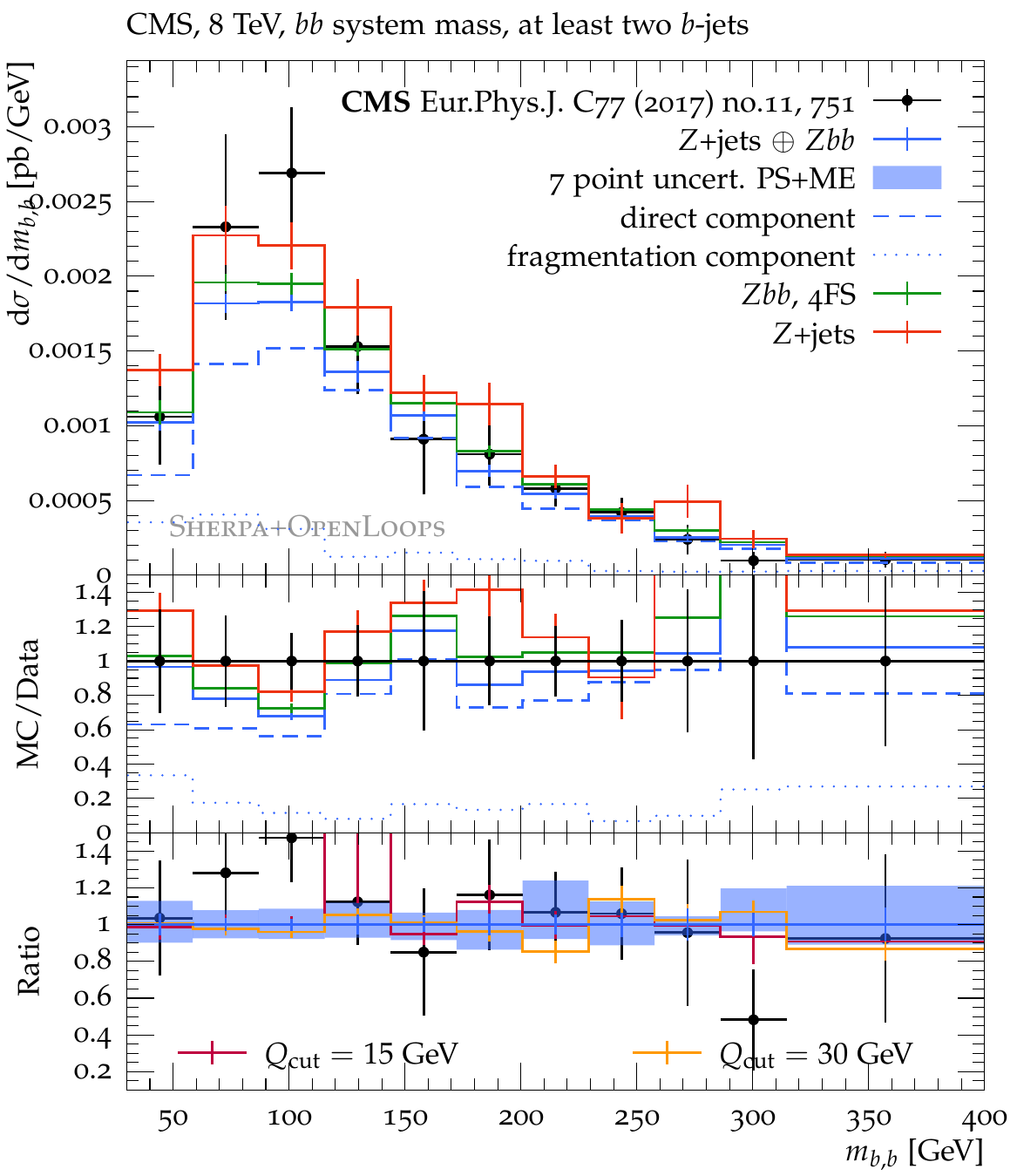}}
  \subfloat[The angular distance of the both leading $b$-jets.]{\includegraphics[width=0.45\textwidth]{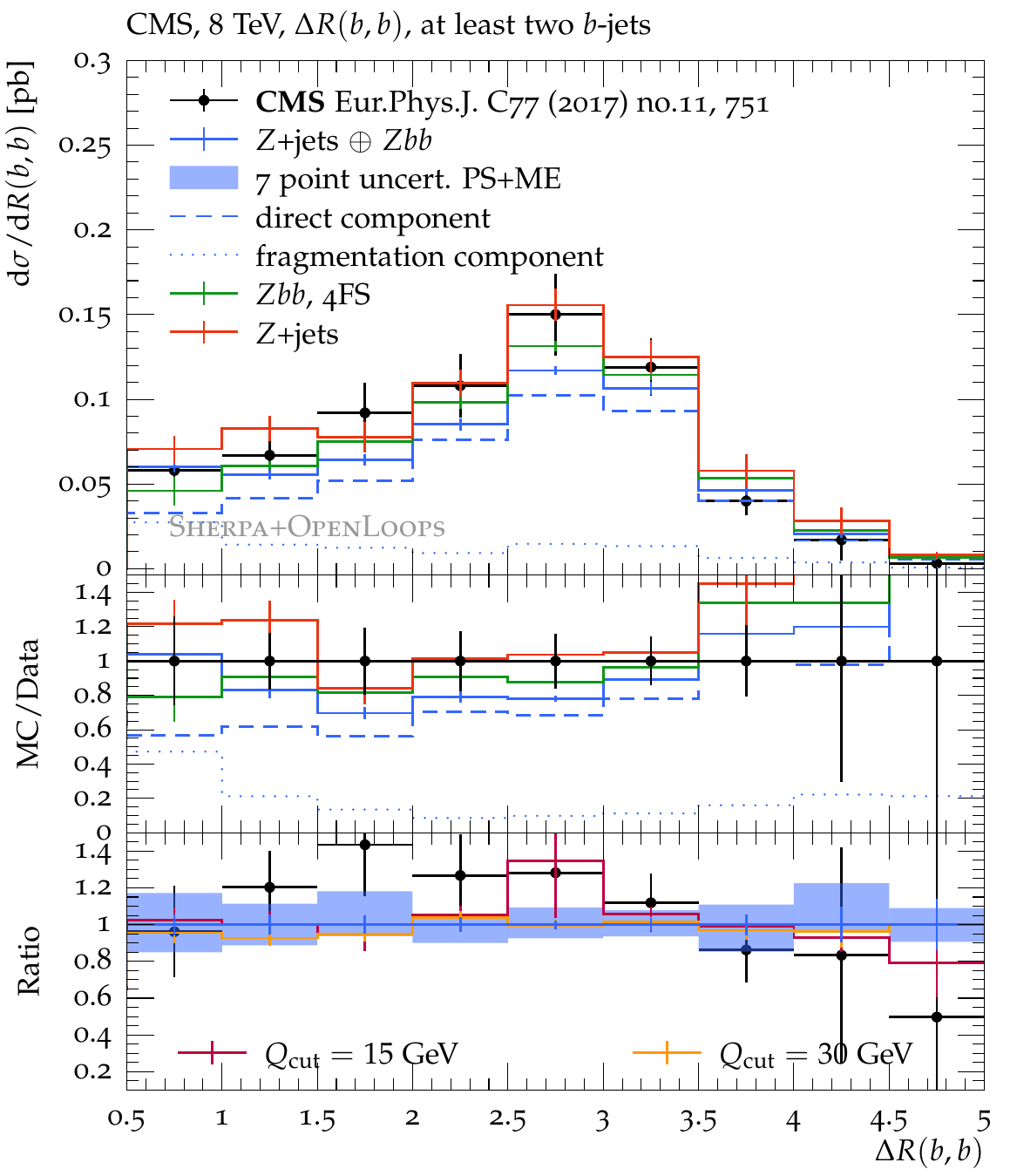}} \\
  \subfloat[The transverse momentum of the $Z$ boson.]{\includegraphics[width=0.45\textwidth]{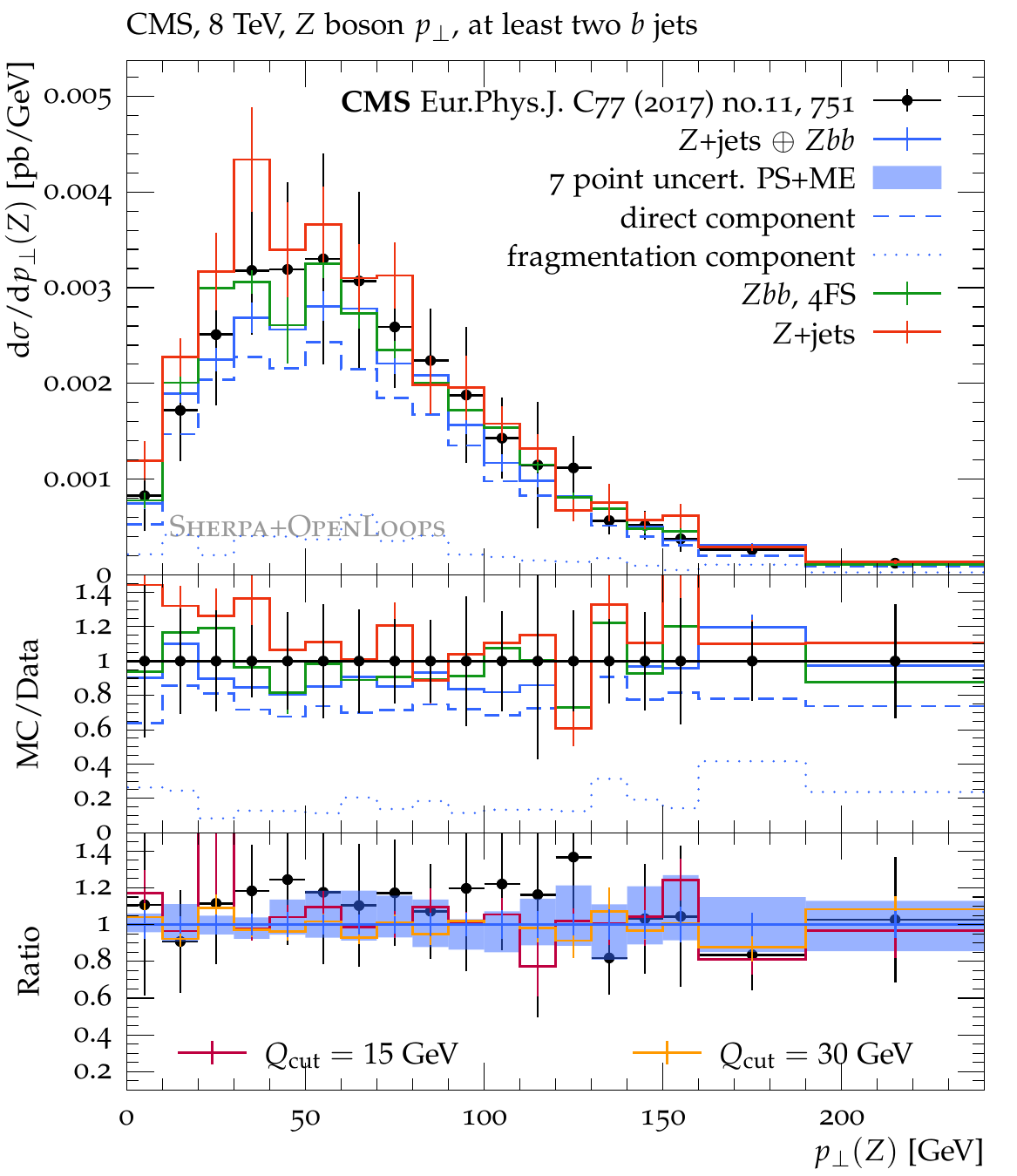}}
  \subfloat[The asymmetry $A_{Zbb}$, defined as $A_{Zbb} = \frac{\mathrm{\Delta}R^\text{max}(Z,b)
		- \mathrm{\Delta}R^\text{min}(Z,b) }{\mathrm{\Delta}R^\text{max}(Z,b) + \mathrm{\Delta}R^\text{min}(Z,b) }$.
		]{\includegraphics[width=0.45\textwidth]{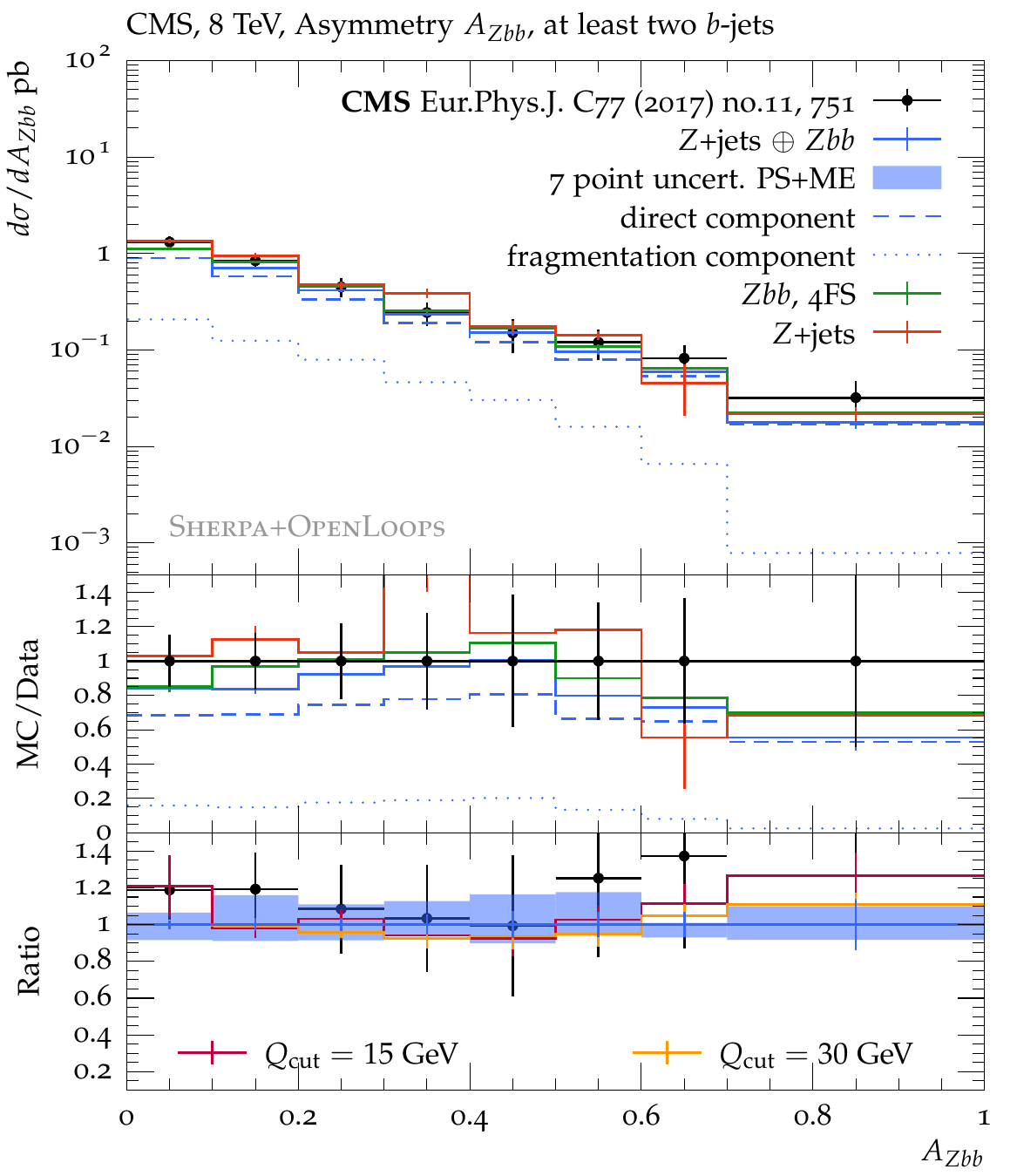}}
  \caption{Comparison of the newly developed algorithm with data from CMS in a phase space region with at least two $b$-jets. 
	    The new fusing prediction is shown in solid blue, its both components
	    are given as dashed and dotted blue lines. The 5FS MEPS@NLO and the 4FS MC@NLO are shown for comparison in red
	    and green. The uncertainties for the fused prediction are shown in the second ratio plot. 
	    They include a simultaneous seven-point variation of $\muf$ and $\mur$ for both, matrix element and parton shower emissions,
	    and a merging cut variation for $Q_\text{cut}=15(30)~\GeV$.} 
  \label{fig:z_two_b}
\end{figure}

For the validation of our newly developed approach in a $Zbb$ region we use
 8~\TeV data taken by CMS~\cite{Khachatryan:2016iob}.
There, the $Z$ boson is reconstructed from either two electrons or two muons in a mass
window between $71~\GeV$ and $111~\GeV$. These leptons are required to have $\pt>20~\GeV$ and
$|\eta|<2.4$, and are dressed with photons within a cone of $\Delta R<0.1$.
Jets are defined by the anti-$k_t$ algorithm with $R=0.5$, $\pt>30~\GeV$ and $|\eta|<2.4$.
Only jets with no overlap ($\Delta R>0.5$) to leptons are taken into account.
$b$-jets are identified by ghost association~\cite{Cacciari:2007fd} and have to pass the same jet cuts as described above.

Again, we compare our newly obtained prediction to the experimental data and to predictions obtained in
the 4FS and 5FS. In addition, we estimate the perturbative uncertainties and the uncertainties related to the merging
scale. 
The former are given by 7-point variations of $\mur$ and $\muf$\footnote{We vary
$\mur$ and $\muf$ independently by factors of $0.5$ and $2$,
excluding variations in opposite directions.} coherently in
the matrix elements and the parton shower. The latter are studied by a
variation of the merging cut to values of $15$ or $30~\GeV$.
The total cross sections for having at least one or at least two $b$-jets are
displayed in Table~\ref{tab:z_b_xs} and differential distributions for one (two) $b$-jet observables are shown in Fig.~\ref{fig:z_one_b} (Fig.~\ref{fig:z_two_b}).

%one b-jet
In the one $b$-jet region, the predicted cross sections of the fused result and the 4FS prediction
 are in good agreement with the data, 
whereas the 5FS prediction exceeds it by 34~\%.
Differential cross sections for several distributions are given in Fig.~\ref{fig:z_one_b}.
In general, the fused prediction is in good agreement with the data and in between the 4FS and 5FS predictions.
Whereas the former slightly undershoots the data, the latter has a significantly larger cross section.
This holds in particular for small transverse momenta of either the $b$-jet or the $Z$ boson or if the $b$-jet
is close to the $Z$ boson.
Both fusing components are equally relevant in all distributions, with a highly non-trivial phase-space dependence of their relative contributions.
While they are relatively flat and equal in the transverse-momentum spectrum
of the leading $b$-jet, the direct component dominates around the peak of $\pt(Z)$.
A different composition is found in the $S_T$ distribution and for $\mathrm{\Delta}\Phi(Z,b)$. At large $S_T$ or 
small $\mathrm{\Delta}\Phi(Z,b)$
the fragmentation contribution takes over and the direct one only contributes with around 20~\%.
At the same time, the 4FS prediction undershoots the high $S_T$ region.
This region is sensitive to a good modeling of multiple hard jets, which can
only be predicted reliably by multi-jet matrix elements.
This is the case in the fusing procedure, where in the fragmentation component the hard emissions are
generated first by matrix elements and $b$ quarks may be produced later on in the shower.
In the 4FS prediction on the other hand, the $b$-quarks are always described by matrix elements and additional
emissions of light partons by the parton shower can form hard jets in the end. Thus,
the modeling of this region with a 4FS prediction becomes very sensitive to the parton shower which is applied outside its region of validity.
The uncertainties related to the merging scale in the fragmentation component and the perturbative 
uncertainties are depicted in the lower ratio-plots in Fig.~\ref{fig:z_one_b} for all distributions. 
The perturbative uncertainties are at the level of 15-20~\% in all observables. 
Results with different merging cuts are all within the perturbative uncertainty band and in good agreement with each other.
The $Q_\text{cut}=15~\GeV$ curve has significantly higher statistical uncertainties,
which is a typical feature for multi-jet merged predictions with very low merging
scales.

% two b-jets
In the two-$b$-jet region, the total predicted cross sections of both the 4FS and 5FS are in good agreement
with the experimentally measured ones. The fused prediction is slightly lower but still matches the data within the uncertainties.
Differential distributions for this region of phase space are given in Fig.~\ref{fig:z_two_b}. Here, both
the 4FS and the 5FS curve are in good agreement with each other and with the experimental data. The fused prediction
follows very closely the 4FS result but has a slightly smaller cross section in some bins.
In all regions of phase space, the direct component gives the dominant contribution of the fused prediction.
Only for $b$-jet pairs which are collimated or have a small invariant mass, the fragmentation component
exceeds the 20~\% threshold, which demonstrates the expected transition towards unresolved one-$b$-jet configurations.
The perturbative uncertainties are reduced in comparison to the one-$b$-jet region. They are at the level of 5~\% 
for low values of $\pt^Z$ and reach up to 15~\% for larger values. Again, all merging cut variations are 
within the perturbative uncertainty band, with the $Q_\text{cut}=15~\GeV$ curve again yielding large statistical
uncertainties in some bins.
It is worth noting that the fused prediction has a significantly smaller statistical uncertainty than the 5FS prediction, although
the fragmentation component was generated with the same number of events. This is expected in all regions of phase space where
the direct component dominates since only a small fraction of 5FS events will yield two $b$-jets.
The direct component profits from its explicit production of heavy-flavor final states and can fill the phase space more efficiently.

\FloatBarrier

\section{Conclusions}

We have presented a novel event generation algorithm to simulate heavy-flavor
associated production in collider experiments. Building upon the established
merging algorithms for multi-jet matrix elements and parton showers, we propose
a technique to include massive matrix elements for heavy-quark production,
effectively leading to an MC simulation in a variable-flavor-number scheme,
which we call fusing.

The overlap between the five- and four-flavor scheme calculations is removed
based on a parton-shower interpretation of the full parton evolution from the hard scale
to the parton shower cut-off. This evolution history is also used to supplement
the massive matrix elements with all higher-order corrections necessary to maintain
the logarithmic accuracy of the multi-jet merged calculation.

Our algorithm allows to combine the advantages of inclusive five-flavor scheme
calculations with the higher precision of four-flavor scheme calculations in regions
of phase space where the bottom quark mass sets a relevant scale.
Such a combined prediction is crucial for heavy-flavor measurements
in LHC experiments, since they will always be affected by the presence of fake
heavy-flavor tagged jets.

The fusing algorithm can be applied at leading order or next-to-leading order QCD.
Its logarithmic accuracy depends on the parton shower used in the merging and
might be extended to NLL in the near future.
The relation to the FONLL method has been established by analytically
identifying the known FONLL matching coefficients within the fused parton
shower expressions.

Using an implementation in the Sherpa event generator, we show a first
application to heavy-flavor production in association with a $Z$ boson. Cross
checks of exclusive observables and a comparison of the results for heavy-flavor
production to experimental data demonstrate the improvement over existing
multi-jet merging algorithms.

\section*{Acknowledgments}
  We are grateful to our colleagues in the \textsc{Atlas} and \textsc{Sherpa}
  collaborations for numerous discussions and support. We thank in particular
  Thomas Gehrmann, Davide Napoletano and Maria Ubiali for discussions on the
  connection to the FONLL method, Marek Sch\"onherr and Enrico Bothmann for
  discussions on the general theoretical background, and Judith Katzy and
  Chris Pollard for fruitful interactions about the application of this
  technique to $t\bar{t}b\bar{b}$ measurements. We are grateful to John Campbell
  for discussions of the method and for his comments on the manuscript.
  This work was supported by the German Research Foundation (DFG)
  under grant No.\ SI 2009/1-1.
  It used resources of the Fermi National Accelerator Laboratory (Fermilab),
  a U.S. Department of Energy, Office of Science, HEP User Facility.
  Fermilab is managed by Fermi Research Alliance, LLC (FRA), acting under
  Contract No. DE--AC02--07CH11359.
  We thank the Center for Information Services and High Performance Computing
  (ZIH) at TU Dresden for generous allocations of computing time.

\appendix
\section{Correspondence between inclusive and exclusive parton evolution}
\label{app:nlo_vs_ps}
In this appendix we summarize the correspondence between the exclusive
parton evolution implemented by parton showers and the underlying inclusive
evolution equations. In the collinear limit, the evolution of parton
densities $f_a(x,t)$ is determined by the DGLAP equations~\cite{
  Gribov:1972ri,Lipatov:1974qm,Dokshitzer:1977sg,Altarelli:1977zs}
\begin{equation}\label{eq:pdf_evolution}
  \frac{{\rm d}\,xf_{a}(x,t)}{{\rm d}\ln t}=
  \sum_{b=q,g}\int_0^1{\rm d}\tau\int_0^1{\rm d} z\,\frac{\alpha_s}{2\pi}
  \big[z\hat{P}_{ba}(z)\big]_+\,\tau f_{b}(\tau,t)\,\delta(x-\tau z)\;.
\end{equation}
In this context, $\hat{P}_{ba}$ are the unregularized DGLAP evolution kernels, which can be expanded
into a power series in the strong coupling. The plus prescription is employed to enforce the 
momentum and flavor sum rules as
\begin{equation}\label{eq:sf_regularization}
  zP_{ba}(z)=\big[z\hat{P}_{ba}(z)\big]_+=\lim\limits_{\eps\to 0}z\hat{P}_{ba}(z,\eps)\;,
\end{equation}
where
\begin{equation}\label{eq:kernels_fineps}
  \hat{P}_{ba}(z,\eps)=\hat{P}_{ba}(z)\,\Theta(1-z-\eps)
    -\delta_{ab}\sum_{c\in\{q,g\}}
    \frac{\Theta(z-1+\eps)}{\eps}
    \int_0^{1-\eps}{\rm d}\zeta\,\zeta\,\hat{P}_{ac}(\zeta)\;.
\end{equation}
For finite $\eps$, the endpoint subtraction in Eq.~\eqref{eq:sf_regularization} 
can be interpreted as the approximate virtual plus unresolved real corrections, 
which are included in the parton shower because the Monte-Carlo algorithm
implements a unitarity constraint~\cite{Jadach:2003bu}. The precise value 
of $\eps$ is determined in terms of an infrared cutoff on the evolution variable,
by means of four-momentum conservation~\cite{Hoeche:2017jsi}. For $0<\eps\ll 1$,
Eq.~\eqref{eq:pdf_evolution} can be written as
\begin{equation}\label{eq:pdf_evolution_constrained}
  \frac{1}{f_{a}(x,t)}\,\frac{{\rm d} f_{a}(x,t)}{{\rm d}\ln t}=
  -\sum_{c=q,g}\int_0^{1-\eps}{\rm d}\zeta\,\zeta\,\frac{\alpha_s}{2\pi}\hat{P}_{ac}(\zeta)\,
  +\sum_{b=q,g}\int_x^{1-\eps}\frac{{\rm d} z}{z}\,
  \frac{\alpha_s}{2\pi}\,\hat{P}_{ba}(z)\,\frac{f_{b}(x/z,t)}{f_{a}(x,t)}\;.
\end{equation}
Using the Sudakov factor of the parton shower
\begin{equation}\label{eq:sudakov}
  \Delta_a(t_0,t)=\exp\bigg\{-\int_{t_0}^{t}\frac{{\rm d} \bar{t}}{\bar{t}}
  \sum_{c=q,g} \int_0^{1-\eps}{\rm d}\zeta\,\zeta\,\frac{\alpha_s}{2\pi}\hat{P}_{ac}(\zeta)\bigg\}
\end{equation}
one can define the generating function for splittings of parton $a$ as
\begin{equation}\label{eq:def_updf}
  \mathcal{F}_a(x,t,\mu^2)=f_a(x,t)\Delta_a(t,\mu^2)\;.
\end{equation}
Equation~\eqref{eq:pdf_evolution_constrained} can then be reduced to
\begin{equation}\label{eq:pdf_evolution_constrained_2}
  \frac{{\rm d}\ln\mathcal{F}_a(x,t,\mu^2)}{{\rm d}\ln t}
  =\sum_{b=q,g}\int_x^{1-\eps}\frac{{\rm d} z}{z}\,
  \frac{\alpha_s}{2\pi}\,\hat{P}_{ba}(z)\,\frac{f_{b}(x/z,t)}{f_{a}(x,t)}\;.
\end{equation}
It is solved using the Markovian Monte-Carlo techniques implemented by parton showers.
Note that Eq.~\eqref{eq:pdf_evolution_constrained_2} is structurally equivalent
to the inclusive result, Eq.~\eqref{eq:pdf_evolution}, which can be written as
\begin{equation}\label{eq:pdf_evolution_2}
  \frac{{\rm d}\ln f_{a}(x,t)}{{\rm d}\ln t}=
  \sum_{b=q,g}\int_x^1\frac{{\rm d} z}{z}\,
  \frac{\alpha_s}{2\pi}\,P_{ba}(z)\,\frac{f_{b}(x/z,t)}{f_{a}(x,t)}\;.
\end{equation}
In the context of our analysis in Sec.~\ref{sec:fonll_nlonll} it is important
to remember this formal correspondence, which can be rephrased as the equivalence
of the regularized DGLAP splitting kernels, $P_{ab}(z)$, and the splitting kernels
$P_{ab}(z,\eps)$ defined in Eq.~\eqref{eq:kernels_fineps} as $\eps\to0$.
While this limit cannot be taken in the parton shower in practice~\cite{Hoeche:2017jsi},
it is a useful theoretical construction to prove that the Monte-Carlo algorithm
and the inclusive parton evolution generate formally identical results to any
logarithmic accuracy that is implemented by both calculations.

\section{Leading and next-to-leading order coefficient functions}
\label{sec:sfs}
The leading-order DGLAP splitting functions used in Sec.~\ref{sec:fonll}
are defined as~\cite{Gribov:1972ri,Dokshitzer:1977sg,Altarelli:1977zs}
\begin{equation}
  \begin{split}
    P_{qq}(z)=&\;C_F\,\frac{1+z^2}{(1-z)_+}+\frac{3}{2}\,C_F\,\delta(1-z)\;,
    &P_{qg}(z)=&\;C_F\,\frac{1+(1-z)^2}{z}\;,\\
    P_{gq}(z)=&\;T_R\left(1-2z(1-z)\right)\;,
    &P_{gg}(z)=&\;2C_A\left(\frac{z}{(1-z)_+}+\frac{1-z}{z}+z(1-z)\right)+\beta_0\,\delta(1-z)\;.
  \end{split}
\end{equation}
The next-to-leading order splitting functions needed to evaluate
Eq.~\eqref{eq:fonll_coefficients} are given by~\cite{Curci:1980uw,
  Furmanski:1980cm,Floratos:1980hk,Floratos:1980hm,Heinrich:1997kv,Ellis:1996nn}
\begin{equation}
  \begin{split}
    &P_{qq}^{S,(1)}(z)=C_FT_R\left[-(1+z)\ln^2z+\left(\frac{8}{3}z^2+5z+1\right)\ln z-\frac{56}{9}z^2+6z-2+\frac{20}{9z}\,\right]\;,\\
    &P_{gq}^{(1)}(z)=C_FT_R\bigg[\,2-\frac{9}{2}z-\left(\frac{1}{2}-2z\right)\ln z-\left(\frac{1}{2}-z\right)\ln^2z+2\ln(1-z)\\
      &\qquad\qquad\qquad\qquad+\bigg(\ln^2\frac{1-z}{z}-2\ln\frac{1-z}{z}-\frac{\pi^2}{3}+5\bigg)\frac{P_{gq}(z)}{T_R}\bigg]\\
    &\qquad+C_AT_R\bigg[\frac{91}{9}+\frac{7}{9}z+\frac{20}{9z}+\left(\frac{68}{3}z-\frac{19}{3}\right)\ln z
      -2\ln(1-z)-(1+4z)\ln^2z+S_2(z)\frac{P_{gq}(-z)}{T_R}\\
      &\qquad\qquad\qquad\quad+\left(-\frac{1}{2}\ln^2z+\frac{22}{3}\ln z-\ln^2(1-z)+2\ln(1-z)
      +\frac{\pi^2}{6}-\frac{109}{9}\right)\frac{P_{gq}(z)}{T_R}\,\bigg]\;.\\
    &P_{qq,Q}^{V,(1)}(z)=C_FT_R\left[-\left(\frac{2}{3}\ln z+\frac{10}{9}\right)\frac{1+z^2}{1-z}-\frac{4}{3}(1-z)\,\right]_+\;,\\
  \end{split}
\end{equation}
The auxiliary function $S_2$ is defined as~\cite{Ellis:1991qj}
\begin{equation}
  \begin{split}
    S_2(z)=\;&-2\,{\rm Li}_2\Big(\frac{1}{1+z}\Big)
    +\frac{1}{2}\log^2 z-\log^2(1-z)+\frac{\pi^2}{6}\;.
  \end{split}
\end{equation}
The $\beta$ function coefficients used in Sec.~\ref{sec:fonll} are
\begin{equation}
  \begin{split}
    \beta_0=&\frac{11}{6}C_A-\frac{2}{3}T_R n_f\;,
    &\beta_{0,b}&=-\frac{2}{3}T_R\;.
  \end{split}
\end{equation}
Note that $n_f$ counts the number of light parton flavors only~\cite{Buza:1995ie}.

\bibliography{journal}
\end{document}